\documentclass[12pt]{article}
\usepackage{amsmath,amssymb,hyperref,graphicx,color,slashed}

\makeatletter
\def\amsbb{\use@mathgroup \M@U \symAMSb}
\makeatother
\usepackage{bbold}
\newcommand{\identity}{\mathbb{1}}

\newcommand{\be}{\begin{equation}}
\newcommand{\ee}{\end{equation}}
\newcommand{\bea}{\begin{eqnarray}}
\newcommand{\eea}{\end{eqnarray}}
\newcommand{\beas}{\begin{eqnarray*}}
\newcommand{\eeas}{\end{eqnarray*}}

\begin{document}
\begin{titlepage}

\enlargethispage{0.5cm}
\vspace*{-1.5cm}

\begin{center}
{\Large Compactification Without Orientation, or} \\[5pt]
{\Large a Topological Scenario for $CP$ Violation}

\vspace{10mm}

\renewcommand\thefootnote{\mbox{$\fnsymbol{footnote}$}}
Brian Greene${}^{1}$\footnote{brian.greene@columbia.edu},
Daniel Kabat${}^{2,3}$\footnote{daniel.kabat@lehman.cuny.edu},
Janna Levin${}^{4}$\footnote{janna@astro.columbia.edu},
Massimo Porrati${}^{5}$\footnote{massimo.porrati@nyu.edu}

\vspace{6mm}

${}^1${\small \sl Departments of Physics and Mathematics, Columbia University} \\
{\small \sl 538 West 120th Street, New York, NY 10027, USA}

\vspace{3mm}

${}^2${\small \sl Department of Physics and Astronomy} \\
{\small \sl Lehman College, City University of New York} \\
{\small \sl 250 Bedford Park Blvd.\ W, Bronx, NY 10468, USA}

\vspace{3mm}

${}^3${\small \sl Graduate School and University Center, City University of New York} \\
{\small \sl  365 Fifth Avenue, New York, NY 10016, USA}

\vspace{3mm}

${}^4${\small \sl Department of Physics and Astronomy} \\
{\small \sl Barnard College of Columbia University} \\
{\small \sl New York, NY 10027, USA}

\vspace{3mm}

${}^5${\small \sl Center for Cosmology and Particle Physics} \\
{\small \sl Department of Physics, New York University} \\
{\small \sl 726 Broadway, New York, NY 10003, USA}

\end{center}

\vspace{10mm}

\noindent
In higher dimensional theories, we often assume that the extra dimensions form an orientable
space, perhaps with singularities.  However, many physical theories are well-defined on non-orientable spaces,
and many spaces are not orientable, so it is reasonable
to explore what happens if the assumption of orientability is relaxed.  Here we consider the simplest example of free 6D theories
compactified on a flat Klein bottle.  We focus on a Dirac fermion in 6D, with boundary conditions that define
${\rm pin}^+$ and ${\rm pin}^-$ structures.  Translation invariance is broken by the boundary conditions, which leads
to sharp features localized near the parity walls (fixed points of the reflection used to construct the Klein bottle).
For a scalar field, there is a position-dependent energy density, peaked near the parity walls.  A Dirac fermion can lead to breaking
of parity, charge conjugation and $CP$ in 3+1 dimensions.  Order parameters for this
breaking are provided by the vacuum expectation values of certain fermion bilinears, again peaked near the parity walls.  As one potential application, these results suggest mechanisms
for $CP$ violation and baryogenesis, enabled by compactification on a Klein bottle.

\end{titlepage}

\setcounter{footnote}{0}
\renewcommand\thefootnote{\mbox{\arabic{footnote}}}

\hrule
\tableofcontents
\bigskip
\hrule

\addtolength{\parskip}{8pt}

\section{Introduction\label{sect:intro}}
For more than a century, physicists have developed theories involving extra dimensions.
In the vast literature on the subject, a dominant assumption is that the extra dimensions form an oriented
manifold.\footnote{For exceptions in the context of string theory, see for example \cite{Dabholkar:1996pc,Freed:2019sco,Cheng:2023owv,Chakrabhavi:2025bfi}.}  This assumption may be more a matter of history and convenience than necessity.  As we will review, to be defined on a non-orientable space, a higher-dimensional theory must have a
parity symmetry, and such theories did not appear to provide promising starting points for phenomenology.  Instead, the general consensus was that the existence of chiral fermions in 4D indicated that the higher dimensional theory must also be chiral \cite{Appelquist:1987nr}.
However, with the modern understanding that chiral fermions can be localized on branes, or on singularities of the compactification,
the case for starting with a chiral theory in higher dimensions becomes less compelling.  There is also the matter of mathematical expediency.  Most physicists are more familiar with orientable manifolds and so have grown accustomed to using the analytical tools these manifolds support.  Although more challenging to study, non-orientable manifolds substantially broaden the arena of compactified theories, potentially offering new insights and features, so it would appear a possibility worth exploring. 

\begin{figure}
\centering{\includegraphics[width=0.284\linewidth]{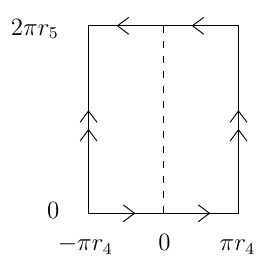}
\hspace{1cm}
\includegraphics[width=0.57\linewidth]{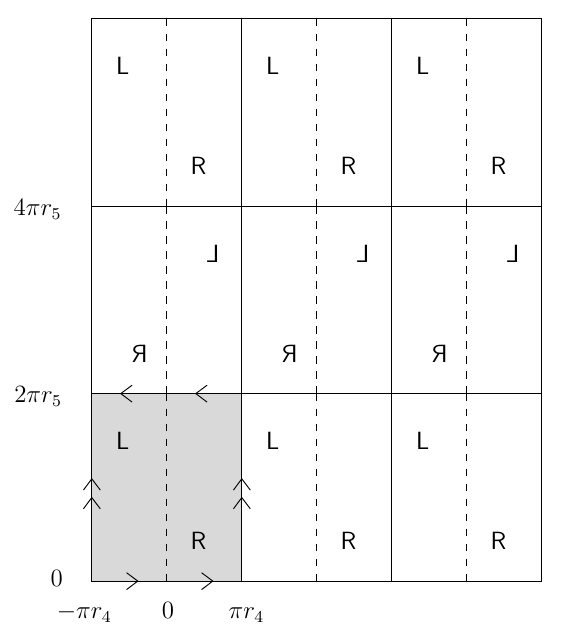}}
\caption{Left: a Klein bottle of size $(2 \pi r_4, 2 \pi r_5)$ built from a single fundamental tile in the $(x^4,x^5)$ plane.  The sides are
identified as shown.  The flip axis at $x^4 = 0$ is indicated by a vertical dashed line.  The identified left and right
sides, at $x^4 = \pm \pi r_4$, form a second flip axis.  Right: the Klein bottle can be lifted to tile the entire
plane.  A fundamental tile, shaded in the lower left corner, and some of its images are shown.  We have put
symbols {\sf L}, {\sf R} on the Klein bottle to help keep track of orientation.  Note that, as an intermediate
step, the Klein bottle can be lifted to a covering torus of size $(2 \pi r_4, 4 \pi r_5)$.}
\label{fig:tiles}
\end{figure}

In the present work, we begin with some background regarding spinors on non-orientable manifolds.  We then explore a simple, prototypical example: a 6D theory compactified to ${\amsbb R}^{3,1}$ on a Klein
bottle.  The Klein bottle is the simplest compact non-orientable space, made by identifying opposite sides of a rectangle,
but with one of the identifications involving a reflection.  The geometry is illustrated in Fig.\ \ref{fig:tiles}.

A theory needs to be parity-symmetric to be defined
on a Klein bottle, and for this reason we focus on a free Dirac fermion in 6D.  We explore possible boundary conditions and
examine the fate of various symmetries.  A curious and somewhat non-intuitive fact is that the symmetries of parity and charge
conjugation in 3+1 dimensions can be broken by the compactification.  We discuss this in section \ref{sect:bc}, where we show
that $P$, $C$ and $CP$ breaking can be detected by the vevs of certain fermion bilinears.  In addition, although
the Klein bottle geometry is locally flat and homogeneous, translation invariance is broken globally by the boundary conditions.  This leads to a
position-dependent Casimir energy density which we evaluate in section \ref{sect:Casimir}.
For completeness, we give the Kaluza-Klein spectrum in appendix \ref{appendix:spectrum}, and for reference,
we work out the 2-point correlators for
spinor and scalar fields in appendix \ref{appendix:correlator}.  In appendix \ref{appendix:pinC} we explore a generalization in which an arbitrary phase is introduced in the boundary conditions, and in appendix \ref{appendix:CR} we touch on the possibility of building
a Klein bottle in which the boundary conditions are twisted by parity combined with charge conjugation.

\subsection{Math background\label{sect:math}}
We begin with a little math background.  For an oriented $d$-dimensional manifold with a Euclidean metric, different coordinate patches are glued together using $SO(d)$ transition functions.  We will be particularly
interested in working with spinors.  To do this, we need to lift the transition functions from $SO(d)$ to ${\rm Spin}(d)$, the double cover of $SO(d)$.\footnote{The additional element generating ${\rm Spin}(d)$
is a $2\pi$ rotation, often denoted $(-1)^F$ since it changes the sign of all spinor fields.}  There may be more than one way to do this, in which case the manifold is said to admit more than one spin structure.

For an unoriented manifold we begin from the group $O(d)$, the extension of $SO(d)$ by an orientation-reversing transformation ${ R}$, and we allow coordinate patches to be glued together by $O(d)$
transformations.  Lifting this structure to spinors is a well-understood but non-trivial task, described for example in \cite{Witten:2015aba}.  The main task is deciding how ${ R}$ acts
on spinors.  One possibility is that ${ R}^2 = 1$, in which case the manifold is said to carry a ${\rm pin}^+$ structure.  Another possibility is that ${ R}^2 = (-1)^F$, so that ${ R}^2$ changes the sign of all spinor fields,
in which case the manifold is said to carry a ${\rm pin}^-$ structure.  It is interesting to ask whether a given unoriented manifold admits ${\rm pin}^+$ structures, ${\rm pin}^-$ structures, both, or neither.  This is discussed, for example, in \cite{BERG_2001,Freed:2016rqq}.  A further possibility is to introduce an arbitrary phase in the definition of $R$, which leads to what is known as a ${\rm pin}_{\amsbb C}$ structure
\cite{Witten:2015aba}.

Since mathematically there is no insurmountable obstruction that prevents defining spinors on at least some unoriented manifolds, we turn now to physical considerations.  Here we encounter a restriction:  for the underlying theory to be well-defined, it must be parity symmetric. That is, it must be invariant under an orientation-reversing transformation $R$.  This would appear to be a serious obstacle, since extensive work on extra-dimensional models in the 1980's \cite{Appelquist:1987nr}
indicated that the most promising approach to obtaining chiral fermions in 4D was to begin with a chiral theory in higher dimensions.  Because such theories are not well defined on non-orientable spaces, historically such choices for the compactified dimensions have received far less attention.  There is, however, a simple way around this obstacle.  We can imagine that the
standard model is localized on a brane, or perhaps on a singularity of the compactification, while the bulk theory has a parity symmetry.

\subsection{Results\label{sect:results}}
As a simple prototype, in what follows we focus on a free Dirac fermion in 6D, compactified to ${\amsbb R}^{3,1}$ on a flat Klein bottle $K_2$.  As a byproduct, we
obtain a few results for a bulk massive scalar.  We introduce boundary conditions built from a reflection that we call ${ R}_4^+$ and ${ R}_4^-$, as well as boundary conditions ${ CR}_4^+$ in which orientation reversal is accompanied by charge conjugation.  These boundary conditions define ${\rm pin}^+$, ${\rm pin}^-$, ${\rm pin}^+$ structures, respectively.  We also consider boundary conditions
that define a ${\rm pin}_{\amsbb C}$ structure, denoted ${ R}_4^\theta$, in which reflection is accompanied by an arbitrary phase.
We examine some physical consequences, finding that a key role is
played by the ``parity walls,'' that is, by the two fixed points (located at $x^4 = 0$ and $x^4 = \pi r_4$) of the reflection $x^4 \rightarrow - x^4$ which is used to
construct the Klein bottle.
\begin{itemize}
\item
In general, translation invariance is broken by the Klein bottle boundary conditions.  For a bulk scalar this leads to a finite, calculable
energy density, sharply peaked near the parity walls, which we compute in section \ref{sect:Casimir}.
\item
With ${ R}_4^+$ boundary conditions, the $U(1)$ global chiral symmetry of a massless fermion $\Psi \rightarrow e^{i \alpha \bar{\Gamma}} \Psi$ is broken, where $\bar{\Gamma}$ is the 6D chirality operator defined in (\ref{6Dchirality}).  The fermion develops a position-dependent
expectation value,
$\langle \bar{\Psi} \bar{\Gamma} \Psi \rangle$, with sharp peaks near the parity walls.  The expectation value breaks parity $P$, defined as a reflection of the spatial coordinates in ${\amsbb R}^{3,1}$, and also
breaks charge conjugation $C$, while leaving the combination $CP$ unbroken.  However, by mixing in transformations that act on the Klein bottle, a rather intricate list of broken and unbroken symmetries
becomes available.  We discuss this in section \ref{sect:R4+}.
\item
With ${ R}_4^-$ boundary conditions, in addition to breaking chiral symmetry $\Psi \rightarrow e^{i \alpha \bar{\Gamma}} \Psi$, the boundary conditions break $CP$.  Again a position-dependent
expectation value develops, peaked at the parity walls, which we denote $\langle \bar{\Psi} \bar{\Gamma}^{(4)} \Psi \rangle$.  Here $\bar{\Gamma}^{(4)}$ is the combination of Dirac matrices defined in
(\ref{Gammabar4}).  The expectation value preserves 4D (but not 6D) Lorentz invariance and serves as
an order parameter for $CP$ breaking.  We explore this in section \ref{sect:R4-}.
\item
${ R}_4^\theta$ boundary conditions interpolate between ${ R}_4^+$ at $\theta = 0$ and ${ R}_4^-$ at $\theta = \pi$.  For generic values of $\theta$ almost all symmetries, including $CP$, are broken.
We explore this generalization in appendix \ref{appendix:pinC}.
\item
With ${ CR}_4^+$ boundary conditions, translation invariance is broken but chiral symmetry is preserved.  Depending on possible phases in the transformations, parity and/or charge conjugation may be broken.
We explore this
briefly in appendix \ref{appendix:CR}.
\end{itemize}
These results have antecedents in the literature.  In particular, working on ${\amsbb R}^{1,1} \times K_2$, the stress tensors for conformally-coupled scalars and massless spin-1/2 fields were obtained in \cite{DeWitt:1979dd}, and fermion bilinears were evaluated in \cite{DeWitt-Morette:1990nrn}.  Early work on pin structures is reviewed in \cite{BERG_2001}.  Our purpose here is to examine the Klein bottle from a contemporary perspective, viewing it as a prototype for non-orientable compactification of a higher-dimensional space, and as a potential source for $CP$ violation in 4D.

\section{Spinor conventions\label{sect:conventions}}
It is convenient to introduce an explicit set of Dirac matrices in 6D as follows.  We begin with a standard chiral basis for the 4D Dirac matrices.
\be
\gamma^0 = \left(\begin{array}{cc} 0 & \identity \\ \identity & 0 \end{array}\right) \qquad
\gamma^i = \left(\begin{array}{cc} 0 & \sigma^i \\ -\sigma^i & 0 \end{array}\right)
\ee
These satisfy
\bea
&& \left\lbrace \gamma^\mu, \gamma^\nu \right\rbrace = 2 \eta^{\mu\nu} \identity \qquad \eta^{\mu\nu} = {\rm diag}(+,-,-,-) \\
\nonumber
&& \hbox{\rm 4D chirality operator $\bar{\gamma} = i \gamma^0 \gamma^1 \gamma^2 \gamma^3 =  \left(\begin{array}{cc} -\identity & 0 \\ 0 & \identity \end{array}\right)$}\,.
\eea
We can build 6D Dirac matrices $\Gamma^M$ as tensor products.
\bea
\nonumber
&& \Gamma^\mu = \gamma^\mu \otimes \sigma^3 = \left(\begin{array}{cc} \gamma^\mu & 0 \\ 0 & -\gamma^\mu \end{array}\right) \qquad \mu = 0,1,2,3,4 \\
&& \Gamma^4 = \identity \otimes i\sigma^1 = \left(\begin{array}{cc} 0 & i\identity \\ i\identity & 0 \end{array}\right) \\
\nonumber
&& \Gamma^5 = \identity \otimes i\sigma^2 = \left(\begin{array}{cc} 0 & \identity \\ -\identity & 0 \end{array}\right)
\eea
These satisfy
\bea
\nonumber
&& \left\lbrace \Gamma^M, \Gamma^N \right\rbrace = 2 \eta^{MN} \qquad \eta^{MN} = {\rm diag}(+,-,-,-,-,-) \\[3pt]
\label{6Dchirality}
&& \hbox{\rm 6D chirality operator $\bar{\Gamma} = \Gamma^0 \Gamma^1 \Gamma^2 \Gamma^3 \Gamma^4 \Gamma^5 = \left(\begin{array}{cc} -\bar{\gamma} & 0 \\ 0 & \bar{\gamma} \end{array}\right)$}\,.
\eea
Note that
\begin{itemize}
\item
$\Gamma^0$ is Hermitian, while all others are anti-Hermitian.
\item
$\Gamma^2$ and $\Gamma^4$ are imaginary, while all others are real.
\item
$\Gamma^0$, $\Gamma^2$, $\Gamma^4$ are symmetric, while $\Gamma^1$, $\Gamma^3$, $\Gamma^5$ are anti-symmetric.
\end{itemize}

With this basis in hand, we can make a few further definitions.  We define the Dirac adjoint by $\bar{\Psi} = \Psi^\dagger \Gamma^0$.  (The relevant property here is that $\Gamma^0 \Gamma^M \Gamma^0 =
\big(\Gamma^M\big)^\dagger$.)  We define a reflection of the $I^{th}$ spatial coordinate via $x \rightarrow x - 2 e_I (e_I \cdot x)$, where $e_I$ is a unit vector, and a corresponding action on spinors, ${ R}_I$, defined by
\be
\label{RI}
{ R}_I \, : \, \Psi(x) \rightarrow \Gamma^I \bar{\Gamma} \Psi\big(x - 2 e_I (e_I \cdot x)\big)\,.
\ee
Finally, charge conjugation acts by
\be
\label{C}
{ C} \, : \, \Psi(x) \rightarrow \Gamma^2 \Gamma^4 \bar{\Gamma} \Psi^*(x)\,.
\ee
Both ${ R}_I$ and ${ C}$ have been defined so that they are symmetries of the massive Dirac Lagrangian in 6D, ${\cal L} = \bar{\Psi} \big(i \Gamma^M \partial_M - m\big)\Psi$.  A possible phase in (\ref{RI}) has been
fixed so that $\big({ R}_I\big)^2 = \identity$.  We will later make use of a modified parity transformation $i { R}_I$ with the property that $\big(i { R}_I\big)^2 = - \identity$.  A possible phase in (\ref{C})
has been fixed as a matter of convention, convenient because the combination $\Gamma^2 \Gamma^4 \bar{\Gamma}$ is real.  Independent of the phase convention used to define $C$, note that applying
$C$ twice to a spinor yields ${ C}^2 \Psi = - \Psi$.\footnote{This makes it impossible to satisfy $\Psi = { C} \Psi$, and is the reason Majorana (self-conjugate)
spinors do not exist in 6D.}

Going forward we will sometimes be cavalier with our notation, so that ${ R}_I$ and $C$ denote the transformations
(\ref{RI}), (\ref{C}) as well as the matrices standing in front of the spinor.  In other words, in addition to
(\ref{RI}) and (\ref{C}), we will use ${ R}_I$ and $C$ to denote
\be
\label{RImatrixCmatrix}
{ R}_I = \Gamma^I \bar{\Gamma} \qquad\quad { C} = \Gamma^2 \Gamma^4 \bar{\Gamma}\,.
\ee

\subsection{Fermion bilinears\label{sect:bilinears}}
For later use, we record the behavior of certain fermion bilinears.  The reason for focusing on these particular bilinears is that, as we will see, they
detect the symmetries that are broken by the Klein bottle boundary conditions.  We will be interested in behavior under reflection ${ R}_I$ and charge conjugation ${ C}$.
We will also be interested in a parity transformation $P$ that changes the sign of the first three spatial coordinates.  We define this by
\be
\label{6Dparity}
{ P} \, : \, \Psi(x) \rightarrow i { R}_1 { R}_2 { R}_3 \Psi(x) = \left(\begin{array}{cc} \gamma^0 & 0 \\ 0 & \gamma^0 \end{array}\right)
\Psi\big(t,-x^1,-x^2,-x^3,x^4,x^5\big)
\ee
(the phase has been chosen so that ${ P}^2 = \identity$).
For future use, we also consider the behavior under $CP$.  Our reason for focusing on these particular transformations is that, as we will see, they provide a convenient set of building
blocks for all of the transformations we wish to consider.

As we mentioned above, the Dirac mass term is invariant under all these transformations, meaning that $\bar{\Psi} \Psi$ is invariant.
The bilinear
\be
\label{PsiGammabarPsi}
i\bar{\Psi} \bar{\Gamma} \Psi\,,
\ee
however, changes sign according to the following table.\footnote{In verifying these signs it helps to use explicit spinor
indices.  See for example \cite{Peskin:1995ev} page 70.} \footnote{The factor of $i$ is included to make $i \bar{\Psi} \bar{\Gamma} \Psi$ real.}

\renewcommand{\arraystretch}{1.2}
\begin{center}
\begin{tabular}{l|cccc}
& $C$ & ${ R}_I$ & $P$ & $CP$ \\
\hline
$\bar{\Psi} \Psi$ & $+$ & $+$ & $+$ & $+$ \\
$i\bar{\Psi} \bar{\Gamma} \Psi$ & $-$ & $-$ & $-$ & $+$
\end{tabular}
\end{center}
It is interesting to contrast this with the behavior of a 4D spinor $\psi$, where we follow the standard definitions for charge conjugation $c$ and parity $p$
in 3+1 dimensions \cite{Peskin:1995ev}.
\bea
\nonumber
{ c} \, : \, \psi(x) & \rightarrow & -i\gamma^2 \psi^*\big(x\big) \\
\label{4Dparity}
{ p} \, : \, \psi(x) & \rightarrow & \gamma^0 \psi\big(t,-x^1,-x^2,-x^3\big)
\eea
In 4D, the scalar and pseudoscalar bilinears transform as
\begin{center}
\begin{tabular}{l|ccc}
& $c$ & $p$ & $cp$ \\
\hline
$\bar{\psi} \psi$ & $+$ & $+$ & $+$ \\
$i\bar{\psi} \bar{\gamma} \psi$ & $+$ & $-$ & $-$
\end{tabular}
\end{center}
Note that in 4D, $i\bar{\psi} \bar{\gamma} \psi$ is odd under $cp$, while in 6D, $i\bar{\Psi} \bar{\Gamma} \Psi$ is $CP$ even.

There is one additional bilinear that will be important later, when we explore the $CP$ breaking associated
with ${ R}_4^-$ boundary conditions.  Define
\be
\label{Gammabar4}
\bar{\Gamma}^{(4)} = i \Gamma^0 \Gamma^1 \Gamma^2 \Gamma^3 = \left(\begin{array}{cc} \bar{\gamma} & 0 \\ 0 & \bar{\gamma} \end{array} \right)
\ee
and consider the bilinear
\be
\label{PsiGamma4Psi}
i \bar{\Psi} \bar{\Gamma}^{(4)} \Psi\,.
\ee
This is not a Lorentz scalar in 6D, but it is invariant under 4D Lorentz transformations of the coordinates $x^\mu$.  It has the same properties under $C$ and $P$ as a pseudoscalar
bilinear in 4D, and it is even under reflections of the coordinates $x^4$, $x^5$.
\begin{center}
\begin{tabular}{l|ccccc}
& $C$ & ${ R}_{1,2,3}$ & ${ R}_{4,5}$ & $P$ & $CP$ \\
\hline
$i\bar{\Psi} \bar{\Gamma}^{(4)} \Psi$ & $+$ & $-$ & $+$ & $-$ & $-$
\end{tabular}
\end{center}

\section{Klein bottle boundary conditions and symmetries\label{sect:bc}}
We are interested in a Dirac fermion in 6D, propagating on ${\amsbb R}^{3,1} \times K_2$ where $K_2$ denotes a Klein bottle.
Starting in ${\amsbb R}^{5,1}$ with coordinates $x^\mu,x^4,x^5$ we make a Klein bottle of size $2 \pi r_4 \times 2 \pi r_5$ by identifying
\bea
\nonumber
&& (x^\mu,x^4,x^5) \approx (x^\mu,x^4 + 2 \pi r_4,x^5) \\
\label{x5period}
&& (x^\mu,x^4,x^5) \approx (x^\mu,-x^4,x^5+2\pi r_5)\,.
\eea
Alternatively we can start by making a covering torus of size $2 \pi r_4 \times 4 \pi r_5$, identifying
\bea
\nonumber
&& (x^4,x^5) \approx (x^4 + 2 \pi r_4,x^5) \\
&& (x^4,x^5) \approx (x^4, x^5 + 4 \pi r_5)\,,
\eea
and then further identify
\be
(x^4,x^5) \approx (- x^4,x^5 + 2 \pi r_5)
\ee
to get a Klein bottle.  To save on writing, we will often denote
\be
\tilde{x} = (x^\mu,-x^4,x^5 + 2 \pi r_5)\,.
\ee

To describe a Dirac fermion on $K_2$, we need to specify boundary conditions.  In what follows, we impose trivial periodicity in $x^4$
\be
\Psi(x^\mu,x^4,x^5) = \Psi(x^\mu,x^4 + 2 \pi r_4, x^5)\,,
\ee
and consider the options (here ${ R}_4$ and $C$ are the matrices defined in (\ref{RImatrixCmatrix}))
\bea
\nonumber
&& { R}_4^+ \, : \,\,\,\, \Psi(x) = { R}_4 \Psi(\tilde{x}) = \Gamma^4 \bar{\Gamma} \Psi(\tilde{x}) \\
\nonumber
&& { R}_4^- \, : \,\,\,\, \Psi(x) = i { R}_4 \Psi(\tilde{x}) = i \Gamma^4 \bar{\Gamma} \Psi(\tilde{x}) \\
\label{CR4condition}
&& { CR}_4^+ \, : \, \Psi(x) = -i { C R}_4 \Psi^*(\tilde{x}) = -i \Gamma^2 \Psi^*(\tilde{x})\,.
\eea
In the last option, a reflection combined with charge conjugation, we introduced $-i$ as a convenient but
somewhat arbitrary choice of phase, selected to make $-i \Gamma^2$ real.  Since
\be
\big({ R}_4\big)^2 = \identity \qquad \big(i { R}_4\big)^2 = - \identity \qquad \big(-i { CR}_4\big)^2 = \identity
\ee
these boundary conditions define ${\rm pin}^+$, ${\rm pin}^-$, ${\rm pin}^+$ structures, respectively.

Note that the boundary conditions we have introduced are not the most general ones compatible with the Klein bottle geometry.  Suppose the fermion is charged under a $U(1)$ gauge symmetry,
with covariant derivative ${\cal D}_M = \partial_M - i q A_M$.  We are free to introduce a flat connection on the Klein bottle, namely a constant gauge field in the $x^5$ direction.
\be
(A_\mu,A_4,A_5) = (0,0,a)
\ee
The gauge field can be eliminated by setting $\Psi = e^{i q a x^5} \Psi'$.  Suppose $\Psi$ satisfies ${ R}_4^+$ boundary conditions.  When expressed in terms of $\Psi'$, the boundary conditions
acquire a phase $e^{i q a 2 \pi r_5}$ on the right-hand side.  We denote this phase by $e^{i \theta/2}$, and call the corresponding boundary conditions ${ R}_4^\theta$.
\be
\label{R4theta}
{ R}_4^\theta \, : \,\,\,\, \Psi(x) = e^{i \theta / 2} { R}_4 \Psi(\tilde{x}) = e^{i \theta / 2} \Gamma^4 \bar{\Gamma} \Psi(\tilde{x})
\ee
This interpolates from ${ R}_4^+$ when $\theta = 0$ to ${ R}_4^-$ when $\theta = \pi$.
This generalization, which is not available for ${ CR}_4^+$, defines what is known as a ${\rm pin}_{\amsbb C}$ structure \cite{Witten:2015aba}.
We explore the consequences in appendix \ref{appendix:pinC}.

In the rest of this section we examine the fate of various symmetries under ${ R}_4^+$ and ${ R}_4^-$ boundary conditions.
A discussion of ${ R}_4^\theta$ boundary conditions may be found in appendix \ref{appendix:pinC}, and a discussion of ${ CR}_4^+$ boundary conditions may be found in appendix \ref{appendix:CR}.  As a preview of coming attractions, the results are summarized
in Table \ref{table:symmetries}.

\begin{table}
\begin{center}
\begin{tabular}{l|cccc}
& $P$ & ${ R}_4$ & ${ R}_5$ & $C$ \\
\hline
${ R}_4^+$ & $-$ & $+$ & $-$ & $-$ \\
${ R}_4^-$ & $-$ & $+$ & $+$ & $+$ \\
${ R}_4^\theta$ & $-$ & $+$ & $- e^{-i \theta}$ & $- e^{-i \theta}$ \\
${ CR}_4^+$ & $-e^{i\phi}$ & $-e^{i\phi}$ & $e^{i\phi}$ & $e^{i\phi}$ 
\end{tabular}
\end{center}
\caption{Compatibility of various discrete symmetries with ${ R}_4^+$, ${ R}_4^-$, ${ R}_4^\theta$ and ${ CR}_4^+$ boundary conditions.  Transformations that preserve the boundary conditions are
indicated with a $+$, transformations that flip the sign of the boundary conditions are indicated with a $-$.  For ${ R}_4^\theta$ the boundary conditions are violated by a $\theta$-dependent phase.  The results
for ${ CR}_4^+$ are from appendix \ref{appendix:CR}, with an arbitrary phase $e^{i \phi / 2}$ inserted in the would-be symmetry transformations (\ref{CR-P}) -- (\ref{CR-C}).\label{table:symmetries}}
\end{table}

A curious fact about the Klein bottle is that parity symmetry in ${\amsbb R}^{3,1}$ can be broken.  This happens for both ${ R}_4^+$ and ${ R}_4^-$ boundary conditions, as can be seen in Table \ref{table:symmetries}.
Before proceeding, the claim that boundary conditions on a Klein bottle can break parity in 3+1 dimensions
might seem a little surprising.  Shouldn't parity in 3+1 dimensions be a geometric transformation of Minkowski space that is independent
of the compactification?  This intuition is correct for bosons, but fails for fermions for elementary but perhaps unfamiliar reasons that we pause to explain.

\begin{figure}
\begin{center}
\includegraphics[width=16cm]{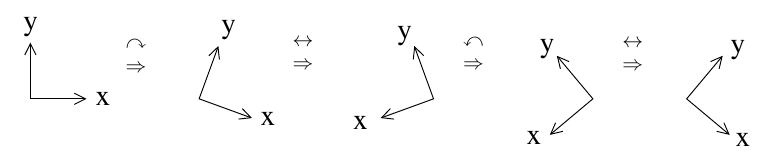}
\includegraphics[width=16cm]{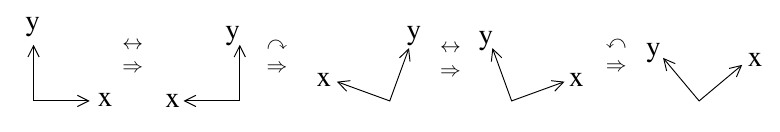}
\end{center}
\caption{Top: the operations that perform
${ R}_x { R}_\theta = { R}_x { D}(\theta) { R}_x { D}(-\theta)$
are equivalent to a rotation ${ D}(-2\theta)$.  Bottom: the operations that perform
$\displaystyle { R}_\theta { R}_x = { D}(\theta) { R}_x { D}(-\theta) { R}_x$
are equivalent to a rotation ${ D}(2\theta)$.\label{fig:rotation}}
\end{figure}

In two spatial dimensions consider a reflection of the $x$ axis
\be
{ R}_x = \left(\begin{array}{cc} -1 & 0 \\ 0 & 1 \end{array}\right)\,.
\ee
Also consider a reflection ${ R}_\theta$ of an axis that has been rotated through an angle $\theta$.  We can build this as
\be
{ R}_\theta = { D}(\theta) { R}_x { D}(-\theta)
\ee
where { D} is a rotation matrix.  The product of two reflections is in the connected component of the rotation group, and as can be seen in Fig.\ \ref{fig:rotation}, we have
\bea
\nonumber
&& { R}_x { R}_\theta = { D}(-2\theta) \\
&& { R}_\theta { R}_x = { D}(2\theta)\,.
\eea
Reflections in different directions in general do not commute, and as $\theta \rightarrow \pi/2$ we find
\bea
\label{RxRy}
&& { R}_x { R}_y = { D}(-\pi) \\
\label{RyRx}
&& { R}_y { R}_x = { D}(\pi)\,.
\eea
Acting on an integer spin field, the rotations (\ref{RxRy}) and (\ref{RyRx}) cannot be distinguished, which leads to the intuition that reflections in orthogonal directions commute.  But acting on a spinor field, they differ by a $2\pi$
rotation, which means that reflections in orthogonal directions anti-commute.  In particular, the reflection on the internal manifold used to build a Klein bottle does not commute with parity in the non-compact dimensions.  In this way,
compactification on a non-orientable manifold can break parity in 3+1.

\subsection{${ R}_4^+$ boundary conditions\label{sect:R4+}}
We implement the Klein bottle identifications (\ref{x5period}) by imposing boundary conditions
\bea
\nonumber
&& \Psi(x^\mu,x^4,x^5) = \Psi(x^\mu,x^4 + 2 \pi r_4,x^5) \\
\label{R4+bc}
&& \Psi(x^\mu,x^4,x^5) = { R}_4 \Psi(x^\mu,-x^4,x^5+2\pi r_5)\,.
\eea
Here
\be
{ R}_4 = \Gamma^4 \bar{\Gamma} = \left(\begin{array}{cc} 0 & i \bar{\gamma} \\ - i \bar{\gamma} & 0 \end{array}\right)
\ee
is the matrix that implements a parity transformation in the $x^4$ direction.  Since $\left({ R}_4\right)^2 = \identity$, this defines a ${\rm pin}^+$ structure
on the Klein bottle.  Going twice around the Klein bottle, the boundary conditions imply
\be
\label{pin+periodic}
\Psi(x^\mu,x^4,x^5) = { R}_4 \Psi(x^\mu,-x^4,x^5+2\pi r_5) = \left({ R}_4\right)^2 \Psi(x^\mu,x^4,x^5 + 4 \pi r_5)\,.
\ee
Since $\left({ R}_4\right)^2 = \identity$, the field is periodic on the covering torus.

\subsubsection{${ R}_4^+$ symmetries\label{sect:R4+symmetry}}
In general, the Klein bottle breaks translation invariance in the $x^4$ direction.  This is simply the statement that if $\Psi(x)$ satisfies the Klein bottle boundary
conditions, then
\be
\Psi'(x) = \Psi(x^\mu,x^4 + a, x^5)
\ee
generically violates them.  Note that there are two unbroken translations, $a = 0$ and $a = \pi r_4$, so it is more accurate to say that translations in $x^4$
are broken to ${\amsbb Z}_2$.

The ${ R}_4^+$ boundary conditions also generically break the chiral symmetry of a massless fermion.  Setting $\Psi'(x) = e^{i \alpha \bar{\Gamma}} \Psi(x)$, and assuming that $\Psi$
satisfies the boundary conditions, we have
\be
\Psi'(x) = e^{i \alpha \bar{\Gamma}} \Psi(x) = e^{i \alpha \bar{\Gamma}} \Gamma^4 \bar{\Gamma} e^{-i \alpha \bar{\Gamma}} \Psi'(\tilde{x}) =
e^{2 i \alpha \bar{\Gamma}} \Gamma^4 \bar{\Gamma} \Psi'(\tilde{x})\,.
\ee
It is more accurate to say that the chiral $U(1)$ symmetry of a massless Dirac field is broken to ${\amsbb Z}_2$, since $\alpha = 0$ and $\alpha = \pi$
preserve the boundary conditions.

Finally, we consider the discrete symmetries $P$, ${ R}_4$, ${ R}_5$, $C$ defined in section \ref{sect:conventions}.  Suppose $\Psi(x)$ satisfies
the Klein bottle boundary conditions, and consider the parity-transformed field
\be
\Psi'(x) = \left(\begin{array}{cc} \gamma^0 & 0 \\ 0 & \gamma^0 \end{array}\right) \Psi(t,-{\bf x},x^4,x^5)\,.
\ee
Since $\Psi$ satisfies the Klein bottle boundary conditions, this implies
\be
\Psi'(x) = \left(\begin{array}{cc} \gamma^0 & 0 \\ 0 & \gamma^0 \end{array}\right) \Gamma^4 \bar{\Gamma} \Psi(t,-{\bf x},-x^4,x^5 + 2 \pi r_5)\,.
\ee
By inverting the parity transformation we can re-write this in terms of $\Psi'$.
\be
\Psi'(x) = \left(\begin{array}{cc} \gamma^0 & 0 \\ 0 & \gamma^0 \end{array}\right) \Gamma^4 \bar{\Gamma} \left(\begin{array}{cc} \gamma^0 & 0 \\ 0 & \gamma^0 \end{array}\right) \Psi'(t,{\bf x},-x^4,x^5 + 2 \pi r_5)
\ee
With a bit of Dirac algebra, this simplifies to
\be
\Psi'(x) = - \Gamma^4 \bar{\Gamma} \Psi'(\tilde{x})\,.
\ee
Because of the sign, if $\Psi(x)$ satisfies the Klein bottle boundary conditions, then $\Psi'(x)$ violates them.

Another instructive example is $C$.  Suppose $\Psi(x)$ satisfies the Klein bottle boundary conditions, and consider the charge conjugate field
\be
\Psi'(x) = \Gamma^2 \Gamma^4 \bar{\Gamma} \Psi^*(x)\,.
\ee
Since $\Psi$ satisfies the Klein bottle boundary conditions, this implies
\be
\Psi'(x) = \Gamma^2 \Gamma^4 \bar{\Gamma} \big(\Gamma^4 \bar{\Gamma}\big)^* \Psi^*(\tilde{x})\,.
\ee
By inverting charge conjugation we can re-write this in terms of $\Psi'$.
\be
\Psi'(x) = \Gamma^2 \Gamma^4 \bar{\Gamma} \big(\Gamma^4 \bar{\Gamma}\big)^* \bar{\Gamma} \Gamma^4 \Gamma^2 (\Psi')^*(\tilde{x})
\ee
With a bit of Dirac algebra, this simplifies to
\be
\Psi'(x) = - \Gamma^4 \bar{\Gamma} \Psi'(\tilde{x})\,.
\ee
Again, if $\Psi(x)$ satisfies the Klein bottle boundary conditions, then $\Psi'(x)$ violates them.

Various other discrete symmetries can be checked following the same steps, such as
\bea
\nonumber
&& { R}_4 : \quad \Psi'(x) = \Gamma^4 \bar{\Gamma} \Psi(x^\mu,-x^4,x^5) \\[2pt]
&& { R}_5: \quad \Psi'(x) = \Gamma^5 \bar{\Gamma} \Psi(x^\mu,x^4,-x^5)\,.
\eea
The results are summarized in  Table \ref{table:symmetries}.\footnote{In checking the behavior under ${ R}_5$, it is important
that, as shown in (\ref{pin+periodic}), the spinor is periodic on the covering torus.}  Note that one can build combinations that preserve the Klein bottle boundary conditions.  In particular ${ PR}_5$,
${ CR}_5$ and $CP$ all preserve the boundary conditions, as do ${ PR}_4{ R}_5$, ${ CR}_4{ R}_5$ and ${ CPR}_4$.  These transformations provide two sets of unbroken symmetries that a 4D observer
could interpret as parity, charge conjugation and $CP$.  On the other hand, some symmetries are broken by the ${ R}_4^+$ boundary conditions.
In particular $P$, $C$ and ${ CPR}_5$ are broken, as are ${ PR}_4$, ${ CR}_4$ and ${ CPR}_4{ R}_5$.  This pattern of broken symmetries, including symmetries that a 4D observer
could interpret as $CP$, plays a role in the scenario for baryogenesis mentioned in section \ref{sect:R4+pheno}.

\subsubsection{${ R}_4^+$ fermion bilinear\label{sect:R4+bilinear}}
\begin{figure}
    \centering{
    \hbox{\includegraphics[width=0.46\linewidth]{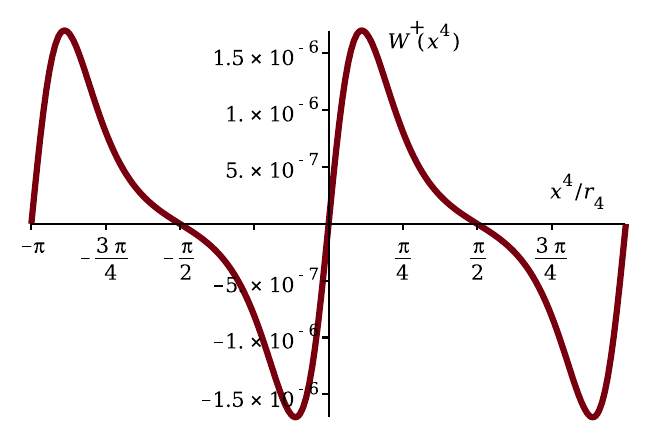} \qquad
    \includegraphics[width=0.46\linewidth]{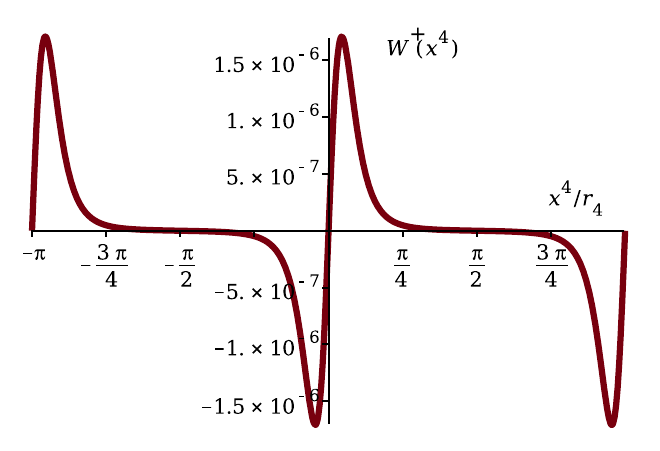}}}
    \caption{The wall function $W^+$ in (\ref{W+}) for a massless field, shown as a function of $x^4/r_4$.  Left: $r_4 = 4$ and $r_5 = 1$. Right: $r_4 = 10$ and $r_5 = 1$.
    When $r_4 \gg r_5$,  the peaks near the origin are well-separated from those near the $x^4=\pm \pi r_4$ edges, and $W^+$ is small in the region between the peaks.}
    \label{fig:wall6D}
\end{figure}

As we have seen, ${ R}_4^+$ boundary conditions break both $P$ and $C$, while preserving the combination
$CP$.  An order parameter for detecting the $P$ and $C$ breaking is the expectation value of the fermion bilinear introduced in (\ref{PsiGammabarPsi}), namely
\be
\langle i \bar{\Psi} \bar{\Gamma} \Psi \rangle\,.
\ee
To evaluate this we write it as (with a minus sign from anti-commuting the fermions)
\be
\label{TracedCorrelator} - i {\rm Tr} \left( \langle \Psi \bar{\Psi} \rangle \bar{\Gamma} \right)\,.
\ee
The Klein bottle correlator with ${ R}_4^+$ boundary conditions, denoted
\be
S_{K_2}^+(y \vert x) = \langle 0 \vert \Psi(y) \bar{\Psi}(x) \vert 0 \rangle \,,
\ee
is evaluated in appendix \ref{appendix:correlator}.
There we show that $S_{K_2}^+$ can be obtained from the correlator on a torus by summing over the orbit of the ${\mathbb Z}_2$ action used to construct the Klein bottle.  This results in an extra term
which we refer to as the Klein image charge.  Using the correlator given in (\ref{SK2+}) and (\ref{ST2pm}) to evaluate (\ref{TracedCorrelator}), the Dirac trace kills most of the terms, since
\be
{\rm Tr} \big(\bar{\Gamma}\big) = {\rm Tr} \big(\Gamma^M \bar{\Gamma}\big) = {\rm Tr} \big(\Gamma^4\bar{\Gamma}\bar{\Gamma}\big) = 0\,.
\ee
The only term that survives comes from the Klein image charge combined with the $\Gamma^4$ term in the torus correlator, which leads to
\be
{\rm Tr} \big(\Gamma^4 \bar{\Gamma} \Gamma^4 \bar{\Gamma}\big) = 8 \,.
\ee
Setting $\epsilon = 0$ since there is no UV divergence, the surviving term is
\bea
\nonumber
\langle i \bar{\Psi} \bar{\Gamma} \Psi \rangle & = & 8 \int_0^\infty {ds \over 16 \pi^2 s^2} e^{-s m^2} \left.{\partial \over
\partial y^4} \right\vert_{y^4 = - x^4} \\
& & {1 \over 2 \pi r_4} \theta_3\Big({y^4 - x^4 \over 2 r_4},e^{-s/r_4^2}\Big)
{1 \over 4 \pi r_5} \theta_3\left.\Big({y^5 - x^5 \over 4 r_5},e^{-s/(2r_5)^2}\Big)\right\vert_{y^5 = x^5 + 2 \pi r_5}
\eea
where $\theta_3$ is a Jacobi theta function.  Some useful properties of theta functions are recorded in appendix \ref{sect:theta}.
Noting that $\theta_3(z,q)$ is even in $z$ and denoting $\theta_3'(z,q) = {\partial \over \partial z} \theta_3(z,q)$, the result can be presented in the form
\be
\label{R4+bilinear}
\langle i \bar{\Psi} \bar{\Gamma} \Psi \rangle = 8 W^+(x^4)
\ee
where
\be
\label{W+}
W^+(x^4) = - {1 \over 2 r_4} \int_0^\infty {ds \over 16 \pi^2 s^2} e^{-s m^2}
{1 \over 2 \pi r_4} \theta_3'\Big({x^4 \over r_4},e^{-s/r_4^2}\Big)
{1 \over 4 \pi r_5} \theta_3\Big({\pi \over 2},e^{-s/(2r_5)^2}\Big)\,.
\ee
This is an odd function of $x^4$, and is periodic with period $\pi r_4$.  It vanishes at the location of the parity walls, i.e.\ at $x^4 = 0$ and $x^4 = \pi r_4$, and
as can be seen in Fig.\ \ref{fig:wall6D}, it has opposite-sign bumps on either side of the walls.  We will refer to $W^+$ as the ${ R}_4^+$ wall function.

\begin{figure}
        \centering
        \includegraphics[width=10cm]{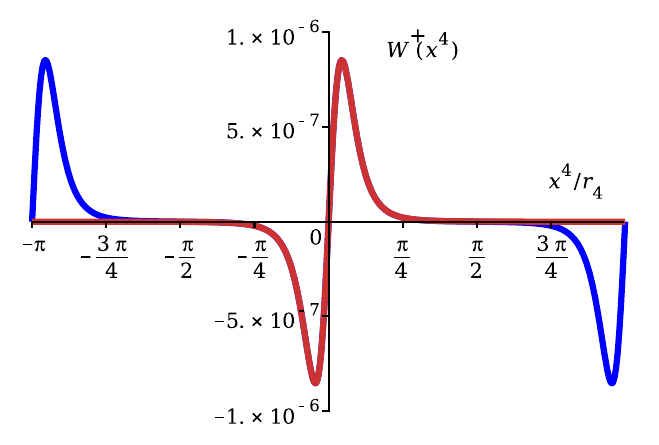}
        \caption{In red, the no-winding term $w_4 = w_5 = 0$ in the sum (\ref{eq:wall6D}) for $W^+(x^4)$. Compare to the blue line, which sums over $w_4 = {-1,0,1}$ with $w_5 = 0$. (The plot is for $r_4 = 10$, $r_5 = 1$.) A second double-bumped wall has appeared around the identified axes $x_4=\pm {\pi}r_4$.}
    \label{fig:cond4Ddisc10}
\end{figure}

The theta function identities (\ref{theta3x4w}), (\ref{theta3x5w}) let us write the result in terms of an image
charge (or winding number) sum, rather than a momentum sum.  The expression simplifies for a massless
field, since the proper-time integral becomes elementary.  This leads to
\be
\label{eq:wall6D}
    W^+(x^4)  =\frac{1}{32\pi^3}
    \sum_{w_4 = - \infty}^\infty \sum_{w_5 = - \infty}^\infty
    \frac{
       x^4 + \pi r_4 w_4
 }{\left[
 (x^4 + \pi r_4 w_4)^2 + \big(\pi r_5 (2w_5+1)\big)^2
\right]^3}\,.
\ee
The winding sum is illustrated in Fig.\ \ref{fig:cond4Ddisc10}.  Note that $W^+(x^4)$ is periodic in $x^4$ with period $\pi r_4$, since shifting $x^4 \rightarrow x^4 + \pi r_4$ can be absorbed by shifting $w_4 \rightarrow w_4 - 1$.  It vanishes at
$x_4 = 0$ because the sum is odd in $w_4$. The periodicity then implies that it vanishes at any multiple of $\pi r_4$.

\subsubsection{A glance at phenomenology\label{sect:R4+pheno}}
With ${ R}_4^+$ boundary conditions, the Klein bottle breaks translation invariance as well as parity
and charge conjugation, while preserving the combination $CP$.  These properties are captured in the expectation
value of the bilinear (\ref{R4+bilinear}).

Note that (\ref{R4+bilinear}) vanishes when averaged over the Klein bottle, so it can be neglected in the traditional
limit of small-volume extra dimensions.  However, the vev provides an interesting background for a braneworld scenario,
in which one imagines a braneworld moving and eventually stabilizing its position on the Klein bottle.  As we show in
subsequent work, coupling a brane fermion to the bulk bilinear can lead to particle production on the brane.  With a few
extra ingredients the Sakharov conditions (including violation of $CP$ symmetry in 4D) can be satisfied, providing a scenario for baryogenesis.

\subsection{${ R}_4^-$ boundary conditions\label{sect:R4-}}
Next we consider ${ R}_4^-$ boundary conditions
\bea
\nonumber
&& \Psi(x^\mu,x^4,x^5) = \Psi(x^\mu,x^4 + 2 \pi r_4,x^5) \\
\label{R4-bc}
&& \Psi(x^\mu,x^4,x^5) = i { R}_4 \Psi(x^\mu,-x^4,x^5+2\pi r_5)
\eea
with
\be
i { R}_4 = i \Gamma^4 \bar{\Gamma} = \left(\begin{array}{cc} 0 & - \bar{\gamma} \\ \bar{\gamma} & 0 \end{array}\right)\,.
\ee
Compared to (\ref{R4+bc}), the only difference is the factor of $i$ in the second line.  This change is more consequential than one might guess, since
as we will see, it leads to some crucial minus signs.  The analysis very much parallels section \ref{sect:R4+}, so we will be brief and only point out where things change.

Since $\left(i{ R}_4\right)^2 = - \identity$, these boundary conditions define a ${\rm pin}^-$ structure on the Klein bottle.  Going twice around the Klein bottle, in place of (\ref{pin+periodic}),
we have
\be
\label{pin-antiperiodic}
\Psi(x^\mu,x^4,x^5) = i { R}_4 \Psi(x^\mu,-x^4,x^5+2\pi r_5) = \left(i { R}_4\right)^2 \Psi(x^\mu,x^4,x^5 + 4 \pi r_5)\,.
\ee
This means the field is anti-periodic on the covering torus.

\subsubsection{${ R}_4^-$ symmetries\label{sect:R4-symmetry}}
Just as in section \ref{sect:R4+}, the ${ R}_4^-$ boundary conditions break both translation invariance and the chiral symmetry of a massless fermion.  More accurately translations in the
$x^4$ direction
\be
\Psi'(x) = \Psi(x^\mu,x^4 + a, x^5)
\ee
are broken to the ${\amsbb Z}_2$ subgroup $a \in \lbrace 0, \pi r_4 \rbrace$, and chiral rotations
\be
\Psi'(x) = e^{i \alpha \bar{\Gamma}} \Psi(x)
\ee
are broken to the ${\amsbb Z}_2$ subgroup $\alpha \in \lbrace 0,\pi \rbrace$.

The fate of the discrete symmetries $P$, ${ R}_4$, ${ R}_5$, $C$ is a little more interesting.  $P$ is still broken, just as in section \ref{sect:R4+},
but let's see what happens to $C$.  Suppose $\Psi(x)$ satisfies the Klein bottle boundary conditions, and consider the charge conjugate field
\be
\Psi'(x) = \Gamma^2 \Gamma^4 \bar{\Gamma} \Psi^*(x)\,.
\ee
Since $\Psi$ satisfies the Klein bottle boundary conditions, this implies (note where the $i$ appears)
\be
\Psi'(x) = \Gamma^2 \Gamma^4 \bar{\Gamma} \big(i \Gamma^4 \bar{\Gamma}\big)^* \Psi^*(x^\mu,-x^4,x^5 + 2 \pi r_5)\,.
\ee
By inverting charge conjugation we can re-write this in terms of $\Psi'$.
\be
\Psi'(x) = \Gamma^2 \Gamma^4 \bar{\Gamma} \big(i \Gamma^4 \bar{\Gamma}\big)^* \bar{\Gamma} \Gamma^4 \Gamma^2 (\Psi')^*(\tilde{x})
\ee
With a bit of Dirac algebra, this simplifies to
\be
\Psi'(x) = + i \Gamma^4 \bar{\Gamma} \Psi'(\tilde{x})\,.
\ee
The $i$ makes a crucial difference: if $\Psi(x)$ satisfies the Klein bottle boundary conditions, so does $\Psi'(x)$.  So $C$ survives as an unbroken symmetry.
Various other discrete symmetries can be checked in the same way.  It turns out that ${ R}_5$ is also unbroken, since the spinor is anti-periodic on the covering
torus.  The results are summarized in Table \ref{table:symmetries}.
Note that with ${ R}_4^-$ boundary conditions, the options for unbroken symmetries in 4D become more limited.  $P$ is broken, but ${ R}_4$, ${ R}_5$ and $C$
all survive.  This means that any transformation involving $P$  is broken and cannot be restored, while any transformation that does not involve $P$ survives as an unbroken symmetry.
In particular $CP$, along with ${ CPR}_4$, ${ CPR}_5$ and ${ CPR}_4{ R}_5$, are all broken by the Klein bottle boundary conditions.

\subsubsection{${ R}_4^-$ fermion bilinear\label{sect:R4-bilinear}}
\begin{figure}
\begin{center}
\includegraphics[width=12cm]{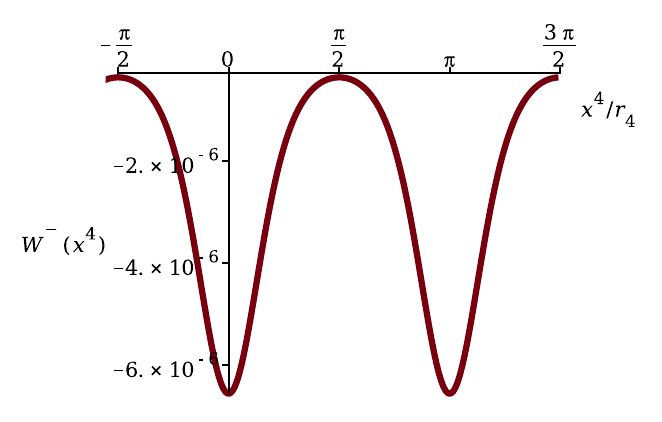}
\end{center}
\caption{A plot of $W^-(x^4)$ vs.\ $x^4/r_4$, for a massless field with $r_4 = 3$ and $r_5 = 1$.\label{fig:W(x)}}
\end{figure}

As we have seen, ${ R}_4^-$ boundary conditions break $P$ and $CP$.  An order parameter for detecting this breaking can be built as a fermion bilinear.  A suitable
bilinear was introduced in (\ref{PsiGamma4Psi}), namely
\be
\label{CPorder}
\langle i \bar{\Psi} \bar{\Gamma}^{(4)} \Psi \rangle\,.
\ee
Recall that $\bar{\Gamma}^{(4)} = i \Gamma^0 \Gamma^1 \Gamma^2 \Gamma^3$, so this expectation value is odd under
$P$ and $CP$ and preserves 4D (but not 6D) Lorentz invariance.  We can evaluate (\ref{CPorder}) by rewriting it as
\be
-i {\rm Tr} \left( \langle \Psi \bar{\Psi} \rangle \bar{\Gamma}^{(4)} \right)
\ee
and using the correlator $S_{K_2}^-$ given in appendix \ref{appendix:correlator}, equations (\ref{SK2-}), (\ref{ST2pm}).
The Dirac trace kills most of the terms, since
\be
{\rm Tr} \, \bar{\Gamma}^{(4)} = {\rm Tr} \left( \Gamma^M \bar{\Gamma}^{(4)} \right) =
{\rm Tr} \left( \bar{\Gamma} \bar{\Gamma}^{(4)} \right) = 0\,.
\ee
The only term that survives comes from the Klein image charge combined with the $\Gamma^5$ term in the torus
propagator.  This leads to
\bea
\nonumber
\langle i \bar{\Psi} \bar{\Gamma}^{(4)} \Psi \rangle & = & {\rm Tr} \left(i \Gamma^4 \bar{\Gamma} \Gamma^5 \bar{\Gamma}^{(4)} \right)
\int_{\epsilon^2}^\infty {ds \over 16 \pi^2 s^2} e^{-s m^2} \left.{\partial \over \partial y^5}\right\vert_{y^5 = x^5 + 2 \pi r_5} \\
& & {1 \over 2 \pi r_4} \theta_3\Big({- 2 x^4 \over 2 r_4},e^{-s/r_4^2}\Big)
{1 \over 4 \pi r_5} \theta_2\Big({y^5 - x^5 \over 4 r_5},e^{-s/(2 r_5)^2}\Big)\,.
\eea
There is no UV divergence so we can set $\epsilon = 0$.  Evaluating the Dirac trace,
the result can be presented as
\be
\label{CPorder2}
\langle i \bar{\Psi} \bar{\Gamma}^{(4)} \Psi \rangle = 8 W^-(x^4)
\ee
where the ${ R}_4^-$ wall function is (note that $\theta_3(z,q)$ is even in $z$)
\be
W^-(x^4) = {1 \over 4r_5} \int_0^\infty {ds \over 16 \pi^2 s^2} e^{-s m^2} {1 \over 2 \pi r_4} \theta_3\Big({x^4 \over r_4},e^{-s/r_4^2}\Big)
{1 \over 4 \pi r_5} \left.{\partial \over \partial z}\right\vert_{z= {\pi \over 2}} \theta_2\Big(z,e^{-s/(2 r_5)^2}\Big)\,.
\ee
As can be seen in Fig.\ \ref{fig:W(x)}, $W^-(x^4)$ is an even periodic function of $x^4$ with period $\pi r_4$.  On the Klein bottle it has two peaks, located at $x^4 = 0$ and at $x^4 = \pi r_4$,
corresponding to the two fixed points of the transformation $x^4 \rightarrow - x^4$.  The wall function can be put in the form of an image sum using the identities (\ref{theta3x4w}), (\ref{theta2x5w}).
This is particularly convenient for a massless field, in which case the result simplifies to
\be
W^-(x^4) = - {1 \over 32 \pi^3} \sum_{w_4,w_5 \in {\amsbb Z}} (-1)^{w_5} {\pi r_5 (2w_5 + 1) \over \left[(x^4 + \pi r_4 w_4)^2 + \big(\pi r_5 (2w_5 + 1)\big)^2\right]^3}\,.
\ee

\subsubsection{A small-volume scenario for $CP$ violation\label{sect:KK-CP}}
With ${ R}_4^-$ boundary conditions, the Klein bottle breaks $CP$.  We could communicate this breaking to the standard model, or at least to a fermion propagating in ${\amsbb R}^{3,1}$,
in a variety of ways, for example, by introducing a coupling to the order parameter (\ref{CPorder2}).  To get an order parameter in 3+1 dimensions, a simple possibility is to extract the zero mode by averaging
(\ref{CPorder2}) over the Klein bottle.
We denote the average over the Klein bottle by
\be
\label{CPorder3}
\langle i \bar{\Psi} \bar{\Gamma}^{(4)} \Psi \rangle_0 = \int_0^{2 \pi r_4} {dx^4 \over 2 \pi r_4} \, \langle i \bar{\Psi} \bar{\Gamma}^{(4)} \Psi \rangle\,.
\ee
Note that, thanks to (\ref{theta3x4n}), extracting the zero mode in this way captures the leading behavior as $r_4 \rightarrow 0$.  So this approach fits with the traditional Kaluza-Klein scenario,
in which we imagine compactifying on a very small Klein bottle.

With the help of
\be
\int_0^{2 \pi r_4} {dx^4 \over 2 \pi r_4} \, \theta_3\Big({x^4 \over r_4},e^{-s/r_4^2}\Big) = 1
\ee
and
\be
\left.{\partial \over \partial z}\right\vert_{z={\pi \over 2}} \theta_2\Big(z,e^{-s/(2 r_5)^2}\Big) = - \left({4 \pi r_5 \over \sqrt{4 \pi s}}\right)^3
2 \sum_{j = 0}^\infty (-1)^j (2j+1) e^{-(4 \pi r_5)^2 (j + {1 \over 2})^2 / 4s}
\ee
the average can be presented more explicitly as
\be
\langle i \bar{\Psi} \bar{\Gamma}^{(4)} \Psi \rangle_0 = - {32 \pi r_5 \over r_4} \int_0^\infty {ds \over (4 \pi s)^{7/2}}
\sum_{j = 0}^\infty (-1)^j (2j+1) e^{-(4 \pi r_5)^2 (j + {1 \over 2})^2 / 4s} e^{-s m^2}\,.
\ee
For a massless field the proper-time integral is elementary and the average simplifies to
\be
\langle i \bar{\Psi} \bar{\Gamma}^{(4)} \Psi \rangle_0 = - {3 \over 16 \pi^7 r_4 (r_5)^4} \sum_{j=0}^\infty {(-1)^j \over (2j+1)^4}
\approx - {6.14 \times 10^{-5} \over r_4 (r_5)^4}\,.
\ee

As a prototype for communicating this breaking to the standard model, we consider a fermion $\psi$ in 4D whose origin
we do not specify, and we introduce a coupling
\be
\label{4fermi}
\int d^4x \, g \, \bar{\psi} \bar{\gamma} \psi \, \langle \bar{\Psi} \bar{\Gamma}^{(4)} \Psi \rangle_0\,.
\ee
This coupling preserves 4D Lorentz invariance.  As mentioned in section \ref{sect:conventions}, the bilinear
$i\bar{\psi} \bar{\gamma} \psi$ is odd under a $cp$ transformation in 4D.  So the interaction (\ref{4fermi}) is invariant under a
simultaneous $cp$ transformation on $\psi$ and $CP$ transformation on $\Psi$.  When $\bar{\Psi} \bar{\Gamma}^{(4)} \Psi$
acquires a vev, it induces a $cp$-violating mass term for $\psi$ in 4D.

\section{Casimir energy and wall tension\label{sect:Casimir}}
In this section we evaluate the Casimir energy for a Klein bottle.  More specifically we evaluate the energy density $\langle T_{00} \rangle$
for scalar and spinor fields, using the two-point correlators obtained in appendix \ref{appendix:correlator}.  Related calculations on the string
worldsheet, using somewhat different methods, may be found in \cite{Polchinski:1998rq,Polchinski:1998rr}.

The correlators have two terms,
one associated with the covering torus and one associated with the Klein image charge.  As a result, the energy density will also have two terms.
There will be a homogeneous term that gives the Casimir energy for the covering torus, and there will be an inhomogeneous term associated with
the parity walls at $x^4 = 0$ and $x^4 = \pi r_4$.

For reasons that will become clear, we begin by evaluating the energy density for a minimally-coupled scalar field, given by
\be
\label{scalarT}
T_{00} = {1 \over 2} \big(\partial_0 \phi\big)^2 + {1 \over 2} \vert \nabla \phi \vert^2 + {1 \over 2} m^2 \phi^2\,.
\ee
We evaluate $\langle T_{00} \rangle$ using the correlators (\ref{DK2}) at coincident points.  One contribution comes from the first term in (\ref{DK2}),
which is the correlator on the covering torus.  Acting on the torus correlator (\ref{DT2}) with the various derivatives appearing in (\ref{scalarT}) and
assembling terms, we find
\be
\langle T_{00} \rangle \vert_{\rm torus} = {1 \over 2} \int_{\epsilon^2}^\infty {ds \over 16 \pi^2 s^2} \left({1 \over s} - {\partial \over \partial s}\right)
{1 \over 2 \pi r_4} \sum_{n_4 \in {\amsbb Z}} {1 \over 4 \pi r_5} \sum_{n_5 \in {\amsbb Z}} e^{-s\left[(n_4 / r_4)^2 + (n_5 / 2 r_5)^2 + m^2\right]}\,.
\ee
This follows from writing the theta functions in the correlator (\ref{DT2}) in the momentum-sum form (\ref{theta3x4n}), (\ref{theta3x5n}).  By switching to the winding-sum
form (\ref{theta3x4w}), (\ref{theta3x5w}) we obtain
\be
\langle T_{00} \rangle \vert_{\rm torus} = {1 \over 2} \int_{\epsilon^2}^\infty {ds \over 16 \pi^2 s^2} \left({1 \over s} - {\partial \over \partial s}\right)
{1 \over \sqrt{4 \pi s}} \sum_{w_4 \in {\amsbb Z}} e^{-(2 \pi r_4)^2 w_4^2 / 4s}
{1 \over \sqrt{4 \pi s}} \sum_{w_5 \in {\amsbb Z}} e^{-(4 \pi r_5)^2 w_5^2 / 4s}
e^{-sm^2}\,.
\ee
The advantage of the winding-sum form is that the UV divergence is isolated in the term with $w_4 = w_5 = 0$.  This term is divergent, but independent
of the radii $r_4$, $r_5$, so it can be absorbed into a renormalization of the cosmological constant.  Suppressing this term we are left with the
standard Casimir energy density on the covering torus.  Setting $\epsilon = 0$ and integrating by parts, the renormalized energy density is
\be
\label{Ttorus}
\langle T_{00} \rangle \vert_{\rm torus} = - \int_0^\infty {ds \over 128 \pi^3 s^4} \sideset{}{'}\sum_{w_4,\,w_5 \in {\amsbb Z}}
e^{-(2 \pi r_4)^2 w_4^2 / 4s}
e^{-(4 \pi r_5)^2 w_5^2 / 4s}
e^{-sm^2}
\ee
where $\sum\nolimits'$ means that $w_4 = w_5 = 0$ is excluded.  For a massless field the result can be written more transparently as
\be
\langle T_{00} \rangle \vert_{\rm torus} = - {1 \over 64 \pi^9} \sideset{}{'}\sum_{w_4,\,w_5 \in {\amsbb Z}} {1 \over \big((r_4 w_4)^2 + (2 r_5 w_5)^2\big)^3}\,.
\ee

Next we turn to the contribution to $\langle T_{00} \rangle$ that comes from the Klein image charge, which we will refer to as $\langle T_{00} \rangle_{\rm wall}$.
This is given by acting with the various derivatives in the stress tensor (\ref{scalarT}) on the second, Klein image charge terms in the correlators (\ref{DK2}).
For a scalar (upper sign) or pseudo-scalar (lower sign) field we find
\bea
\nonumber
\langle T_{00} \rangle \vert_{\rm wall} & = & \pm {1 \over 2} \int_0^\infty {ds \over 16 \pi^2 s^2}
{1 \over 2 \pi r_4} \sum_{n_4 \in {\amsbb Z}} {1 \over 4 \pi r_5} \sum_{n_5 \in {\amsbb Z}} 
\left({1 \over s} - 2 \Big({n_4 \over r_4}\Big)^2 + \Big({n_4 \over r_4}\Big)^2 + \Big({n_5 \over 2 r_5}\Big)^2 + m^2 \right) \\
\label{Twall}
&& 
e^{-i n_4 2 x^4 / r_4} (-1)^{n_5} e^{-s\left[(n_4 / r_4)^2 + (n_5 / 2 r_5)^2 + m^2\right]}\,.
\eea
There is no UV divergence so we have set $\epsilon = 0$.  We have written the factor in parentheses in (\ref{Twall}) in a somewhat peculiar form,
because the combination $(n_4/r_4)^2 + (n_5 / 2 r_5)^2 + m^2$ can be traded for $- {\partial \over \partial s}$ and integrated by parts.
This leads to
\bea
\nonumber
\langle T_{00} \rangle \vert_{\rm wall} & = & \mp {1 \over 2} \int_0^\infty {ds \over 16 \pi^2 s^2}
{1 \over 2 \pi r_4} \sum_{n_4 \in {\amsbb Z}} {1 \over 4 \pi r_5} \sum_{n_5 \in {\amsbb Z}} 
\left({1 \over s} + 2 \Big({n_4 \over r_4}\Big)^2\right) \\
\label{Twall2}
&& 
e^{-i n_4 2 x^4 / r_4} (-1)^{n_5} e^{-s\left[(n_4 / r_4)^2 + (n_5 / 2 r_5)^2 + m^2\right]}\,.
\eea
The sum over $n_4$ can be expressed in terms of $\theta_3$, see (\ref{theta3x4n}), while the sum over $n_5$ is most conveniently expressed
in terms of
\be
{1 \over 4 \pi r_5} \theta_3\Big({\pi \over 2},e^{-s/(2r_5)^2}\Big) = {1 \over 4 \pi r_5} \sum_{n_5 \in {\amsbb Z}} (-1)^{n_5} e^{-s n_5^2 / (2r_5)^2}\,.
\ee
This leads to
\be
\label{Twall3}
\langle T_{00} \rangle \vert_{\rm wall} = \mp {1 \over 2} \int_0^\infty {ds \over 16 \pi^2 s^2}
\left[\Big({1 \over s} - 2 {\partial \over \partial s}\Big) {1 \over 2 \pi r_4} \theta_3\Big({x^4 \over r_4},e^{-s/r_4^2}\Big)\right]
{1 \over 4 \pi r_5} \theta_3\Big({\pi \over 2},e^{-s/(2r_5)^2}\Big) e^{-sm^2}\,.
\ee
Note that the derivative ${\partial \over \partial s}$ only acts inside the square brackets, i.e.\ it only acts on $\theta_3\big({x^4 \over R^4},e^{-s/r_4^2}\big)$.

\begin{figure}
\begin{center}
\includegraphics[width=11cm]{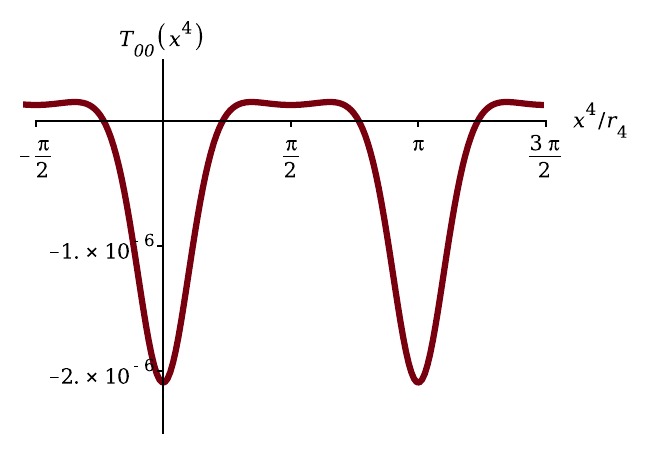}
\end{center}
\caption{The energy density $\langle T_{00} \rangle = \langle T_{00} \rangle \vert_{\rm torus} + \langle T_{00} \rangle \vert_{\rm wall}$
as a function of $x^4/r_4$.  The plot is for a massless scalar (not pseudoscalar) field with $r_4 = 3$ and $r_5 = 1$ in arbitrary units.
(The choice of units sets the scale for $\langle T_{00} \rangle$.)  For these parameters, the Casimir energy density for the covering torus
is small, $\langle T_{00} \rangle \vert_{\rm torus} = - 1.19 \times 10^{-9}$.\label{fig:T00(x)}}
\end{figure}

The total energy density on a Klein bottle is given by the sum of the torus and wall contributions, (\ref{Ttorus}) plus (\ref{Twall3}).  For a scalar
field, the result is shown in Fig.\ \ref{fig:T00(x)}.  One can clearly see a localized energy density associated with the parity walls at $x^4 =0$
and $x^4 = \pi r_4$.  It is worth noting that numerically, it is the wall that makes the dominant contribution.  This can be understood from
the image charge representation of the correlator discussed in appendix \ref{sect:image}.  The first Klein image charge is displaced a distance $2 \pi r_5$, while the first torus image charge is displaced a distance $4 \pi r_5$.  Since the correlator falls off rapidly with distance, the wall contribution dominates.

\begin{figure}
\begin{center}
\includegraphics[width=11cm]{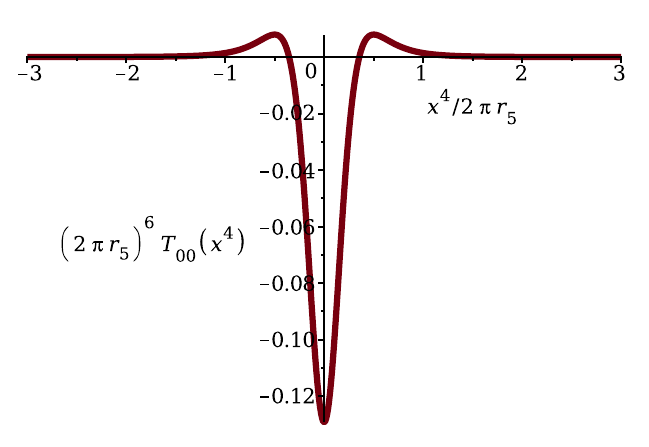}
\end{center}
\caption{Energy density for a massless scalar (not pseudoscalar) field.  The plot shows $\langle T_{00} \rangle \vert_{\hbox{\small single wall}}$ as a function of $x^4$, obtained by
sending $r_4 \rightarrow \infty$ in (\ref{Twall3}).  $r_5$ then sets the scale, so energy density is measured in units of $(2 \pi r_5)^6$ and
position is measured in units of $2 \pi r_5$.\label{fig:wall_energy}}
\end{figure}

It is instructive to study the wall contribution to the energy density in a little more detail.
Let's send $r_4 \rightarrow \infty$ in (\ref{Twall3}) to isolate the contribution of a single wall at $x^4 = 0$.
To do this, we perform a modular transformation on $\theta_3$ using (\ref{theta3x4w}).
This gives
\be
\label{Tsingle-wall}
\langle T_{00} \rangle \vert_{\hbox{\small single wall}} = \mp {1 \over 2} \int_0^\infty {ds \over 16 \pi^2 s^2}
\left[\Big({1 \over s} - 2 {\partial \over \partial s}\Big) {1 \over \sqrt{4 \pi s}} e^{-(x^4)^2/s}\right]
{1 \over 4 \pi r_5} \theta_3\Big({\pi \over 2},e^{-s/(2r_5)^2}\Big) e^{-sm^2}\,.
\ee
By rescaling $s \rightarrow s r_5^2$ one sees that for a massless
field the energy density scales like $1/(r_5)^6$, as required on dimensional grounds.  It should be thought of as a modification to the
bulk Casimir energy density that is associated with the parity walls.
For a massless field, the energy density can be written more transparently as
\be
\langle T_{00} \rangle \vert_{\hbox{\small single wall}} = \mp {1 \over 32 \pi^3 (2 \pi r_5)^6} \sum_{w_5 \in {\amsbb Z}} {(w_5 + 1/2)^2 - 2 (x^4 / 2 \pi r_5)^2 \over
\big((w_5 + 1/2)^2 + (x^4 / 2\pi r_5)^2\big)^4}
\ee
The energy density for a single wall is plotted in Fig.\ \ref{fig:wall_energy}.  

Finally, we turn to the energy density for a bulk fermion, given by
\be
\label{Tfermion}
T_{00} = {i \over 2} \bar{\Psi} \Gamma_0 \partial_0 \Psi - {i \over 2} \partial_0 \bar{\Psi} \Gamma_0 \Psi\,.
\ee
We have delayed considering this case, for the following reason:
the Klein image charge does not contribute to $\langle T_{00} \rangle$.  In other words, a free minimally-coupled bulk fermion does not contribute
to the energy density associated with the parity walls.  This can be shown directly, by examining the Dirac traces
in $\langle T_{00} \rangle$ that involve the Klein image charge, and noting that they all vanish.  This result is special to minimal coupling.  A non-minimal
coupling to curvature, such as $\mathcal{R} \bar{\Psi} \bar{\Gamma} \Psi$ where $\mathcal{R}$ is the scalar curvature, would lead to an energy density localized near the parity walls.

Although it does not contribute to the wall energy, a fermion does contribute to the Casimir energy for the covering torus.  Using the torus correlators
(\ref{ST2+}), (\ref{ST2-}) to evaluate the expectation value of the energy density for ${ R}_4^\pm$ boundary conditions,
a straightforward calculation gives
\be
\label{Ttorus_fermion}
\langle T_{00} \rangle \vert_{\rm torus}^{{ R}_4^\pm} = \int_0^\infty {ds \over 16 \pi^3 s^4} \sideset{}{'}\sum_{w_4,\,w_5 \in {\amsbb Z}}
\left\lbrace {1 \atop (-1)^{w_5}} \right\rbrace
e^{-(2 \pi r_4)^2 w_4^2 / 4s}
e^{-(4 \pi r_5)^2 w_5^2 / 4s}
e^{-sm^2}
\ee
where $\sum\nolimits'$ means that $w_4 = w_5 = 0$ is excluded.  The factor in curly braces stands for $1$ with ${ R}_4^+$
boundary conditions, and for $(-1)^{w_5}$ with ${ R}_4^-$.  Note that, as expected, the torus energy density for fermions
with ${ R}_4^+$ boundary conditions is $(-8)$ times the scalar energy density (\ref{Ttorus}).

\section{Conclusions\label{sect:conclusion}}
Compactification on non-orientable spaces appears to be an under-explored part of the string landscape.  Granted, the bulk theory needs an appropriate
parity symmetry to be defined on a non-orientable space.  However, with the understanding that chiral fermions in 4D can arise from branes or singularities
of the compactification, this underlying parity symmetry is not a major phenomenological obstacle.  Non-orientable compactifications of M-theory have been shown to be
theoretically consistent \cite{Freed:2019sco}, and there are F-theory constructions of non-orientable IIA string compactifications \cite{Cheng:2023owv}.

With this motivation, we explored the consequences of non-orientable compactification in the simple setting of a free Dirac fermion on ${\amsbb R}^{3,1} \times K_2$.
The main surprise was that compactification on $K_2$ could lead to $CP$ breaking in 3+1 dimensions.  The breaking was entirely due to the boundary conditions used to define
the Klein bottle, which meant the $CP$ violating effects were finite and calculable.  The notion of symmetry breaking by boundary conditions is familiar in settings such as the
Hosotani mechanism for breaking gauge symmetry \cite{Hosotani:1988bm} and the Sherk-Schwarz mechanism for breaking supersymmetry \cite{Scherk:1978ta}.

As future directions, it would be interesting to explore the phenomenology and cosmology associated with non-orientable compactification.  Can one build a realistic model for $CP$ violation
in the standard model?  Can one develop a realistic scenario for $CP$ violation and baryogenesis in cosmology?  It would also be interesting to explore more formal questions.  Can one systematically
study the statistics of string vacua associated with non-orientable compactifications?

\bigskip
\goodbreak
\centerline{\bf Acknowledgements}
\noindent
BG is supported in part by DOE award DE-SC0011941. DK is supported by U.S.\ National Science Foundation grant PHY-2412480.  JL is supported in part by the Tow Foundation.
MP is supported in part by NSF grant PHY-2210349.

\appendix
\section{Kaluza-Klein spectrum\label{appendix:spectrum}}
In this section we work out the Kaluza-Klein spectrum on ${\amsbb R}^{3,1} \times K_2$ with ${ R}_4^+$
and ${ R}_4^-$ boundary conditions.  Although the results are not necessary for the remainder of the paper, obtaining the spectrum
and modes makes the physical content clear and gives insight into the behavior under discrete symmetries.

There is one novel feature we will flag in advance.  For fermions, the usual Kaluza-Klein ansatz is a tensor product $\Psi = \psi(x) \otimes \chi(y)$ of a spinor $\psi(x)$ in 4D with a spinor $\chi(y)$
on the internal manifold.  This ansatz can be used to describe modes on the covering torus, as in (\ref{CoveringModes1}), (\ref{CoveringModes2}).  However, the Klein bottle boundary conditions require
a particular linear combination of torus modes, defined in (\ref{KleinSuperposition}).  The linear combination is not a tensor product, and in this way the usual Kaluza-Klein ansatz fails on the Klein bottle.
This is true even for the zero mode (\ref{pin+zeromode}).  In this way, the Klein bottle requires that the internal fermionic modes $\chi$ are intrinsically entangled with the spacetime fermions $\psi$.

\subsection{${ R}_4^+$ spectrum\label{sect:pin+KK}}
We begin by summarizing the results.  With the boundary conditions (\ref{R4+bc}) the Kaluza-Klein tower is labeled by momenta
\be
k_4 = {n_4 \over r_4} \qquad k_5 = {n_5 \over 2 r_5} \qquad n_4,n_5 \in {\amsbb Z}\,.
\ee
For each Kaluza-Klein momentum there is a Dirac fermion $\psi(x)$ in 4D, with mass
\be
m = \sqrt{k_4^2 + k_5^2}\,.
\ee
We define a phase $\alpha$ by $k_5 + i k_4 = m e^{-i \alpha}$.  Then the 4D field lifts to a solution to the 6D Dirac equation
\bea
\nonumber
&& \Psi(x,x^4,x^5) = \left\lbrace\begin{array}{l}
\left(\begin{array}{c} \psi(x) \\ -i \bar{\gamma} \psi(x) \end{array}\right) \qquad \hbox{\rm for $n_4 = n_5 = 0$} \\[20pt]
\left(\begin{array}{c} \psi(x) \\ e^{i \alpha} \psi(x) \end{array}\right) e^{i k_4 x^4} e^{i k_5 x^5}
+ (-1)^{n_5} \left(\begin{array}{c} e^{i \alpha} i \bar{\gamma} \psi(x) \\ - i \bar{\gamma} \psi(x) \end{array}\right) e^{-i k_4 x^4} e^{i k_5 x^5}
\end{array}\right. \\
\label{pin+modes}
&& \hspace{6.6cm} \hbox{\rm for $(n_4,n_5) \not= (0,0)$.}
\eea

We now show how these results are obtained.  The field is periodic on the covering torus, so we can expand in momentum modes
\be
\Psi(x,x^4,x^5) = \left(\begin{array}{l}\chi_1(x) \\ \chi_2(x) \end{array}\right) e^{i k_4 x^4} e^{i k_5 x^5}
\ee
with
\be
k_4 = {n_4 \over r_4} \qquad k_5 = {n_5 \over 2 r_5} \qquad n_4,n_5 \in {\amsbb Z}\,.
\ee
With the definition $k_5 + i k_4 = m e^{-i \alpha}$ the massless Dirac equation in 6D, $i \Gamma^\mu \partial_\mu \Psi = 0$, becomes
\be
\label{6DDirac}
\left(\begin{array}{cc} i \gamma^\mu \partial_\mu & - m e^{-i \alpha} \\[3pt] m e^{i \alpha} & -i \gamma^\mu \partial_\mu \end{array}\right)
\left(\begin{array}{c} \chi_1(x) \\ \chi_2(x) \end{array}\right) = 0\,.
\ee

A special case is $k_4 = k_5 = 0$, for which $m = 0$ and $\alpha$ is ambiguous.  In this case $\chi_1$ and $\chi_2$ separately obey the massless Dirac equation in 4D.  So on the covering torus
we have a pair of Kaluza-Klein zero modes, which can be assembled into a 6D field
\be
\Psi(x,x^4,x^5) = \left(\begin{array}{c} \chi_1(x) \\ \chi_2(x) \end{array}\right) \quad {\rm with} \quad i \gamma^\mu \partial_\mu \chi_1(x) = 0, \quad i \gamma^\mu \partial_\mu \chi_2(x) = 0\,.
\ee
We still have to impose the Klein bottle boundary conditions (\ref{R4+bc}), which require
\be
\left(\begin{array}{c} \chi_1(x) \\ \chi_2(x) \end{array}\right) = \left(\begin{array}{cc} 0 & i \bar{\gamma} \\ - i \bar{\gamma} & 0 \end{array}\right)
\left(\begin{array}{c} \chi_1(x) \\ \chi_2(x) \end{array}\right)\,.
\ee
This just sets $\chi_2 = -i \bar{\gamma} \chi_1$.  So on the Klein bottle there is a single Kaluza-Klein zero mode,
\be
\Psi(x,x^4,x^5) = \left(\begin{array}{c} \psi(x) \\ -i \bar{\gamma} \psi(x) \end{array}\right) \quad {\rm with} \quad i \gamma^\mu \partial_\mu \psi(x) = 0\,.
\ee
We can reproduce this result using a different approach that will be useful below.  Start from a solution on the covering torus
\be
\Psi_1(x,x^4,x^5) = \left(\begin{array}{c} \psi_1(x) \\ 0 \end{array}\right)\,.
\ee
Define the Klein-transformed field
\be
\Psi_1^K(x,x^4,x^5) = \Gamma^4 \bar{\Gamma} \Psi_1(x,-x^4,x^5 + 2 \pi r_5) = \left(\begin{array}{c} 0 \\ -i \bar{\gamma} \psi_1(x) \end{array}\right)\,.
\ee
Then the combination
\be
\label{pin+zeromode}
\hat{\Psi}_1 = \Psi_1 + \Psi_1^K = \left(\begin{array}{c} \psi_1(x) \\ -i \bar{\gamma} \psi_1(x) \end{array}\right)
\ee
satisfies the Klein bottle boundary conditions.\footnote{This is a simple example of building an invariant by averaging over a group of transformations.
It is important that $\Psi^K$ satisfies the Dirac equation and that $K^2 = 1$.  Note that while $\Psi + \Psi^K$ satisfies the Klein bottle boundary conditions, the other linear
combination $\Psi - \Psi^K$ violates them.}  Starting from the other mode on the covering torus $\Psi_2 = \left({0 \atop \psi_2(x)}\right)$ leads
to $\Psi_2 + \Psi_2^K = \left({i \bar{\gamma} \psi_2 \atop \psi_2}\right)$, however this solution is redundant.  To see this, note that $\psi_2 \rightarrow -i \bar{\gamma} \psi_2$
is a symmetry of the massless Dirac equation, and upon making this replacement, one reproduces (\ref{pin+zeromode}).

To obtain the non-zero modes, we start from a solution to the massive Dirac equation in 4D, $i \gamma^\mu \partial_\mu \psi_1(x) = m \psi_1(x)$.  We can build a solution to the 6D Dirac equation on the covering torus as
\be
\label{CoveringModes1}
\Psi_1(x,x^4,x^5) = \left(\begin{array}{c} \psi_1(x) \\ e^{i \alpha} \psi_1(x) \end{array}\right) e^{i k_4 x^4} e^{i k_5 x^5}\,.
\ee
Starting from another solution to the massive Dirac equation in 4D, $i \gamma^\mu \partial_\mu \psi_2(x) = m \psi_2(x)$, we can build a linearly-independent
solution to the Dirac equation on the covering torus as\footnote{This solves the 6D Dirac equation because $\bar{\gamma} \psi_2$ satisfies
$i \gamma^\mu \partial_\mu \bar{\gamma} \psi_2(x) = - m \bar{\gamma} \psi_2(x)$.}
\be
\label{CoveringModes2}
\Psi_2(x,x^4,x^5) = \left(\begin{array}{c} \bar{\gamma} \psi_2(x) \\ - e^{i \alpha} \bar{\gamma} \psi_2(x) \end{array}\right) e^{i k_4 x^4} e^{i k_5 x^5}\,.
\ee
On the covering torus, for every non-zero Kaluza-Klein momentum $(k_4,k_5)$, we obtain a pair $\psi_1$, $\psi_2$ of massive Dirac fields in 4D.

Next we impose the Klein bottle boundary conditions (\ref{R4+bc}), which we write in the form
\be
\Psi(x) = \Psi^K(x)
\ee
where the Klein-transformed field is
\be
\Psi^K(x,x^4,x^5) = \Gamma^4 \bar{\Gamma} \Psi(x,-x^4,x^5 + 2 \pi r_5)\,.
\ee
It is easy to write down linear combinations of the modes (\ref{CoveringModes1}), (\ref{CoveringModes2}) which satisfy the Klein bottle boundary conditions.
\be
\label{KleinSuperposition}
\hat{\Psi}(x,x^4,x^5) = \Psi(x,x^4,x^5) + \Psi^K(x,x^4,x^5)
\ee
Starting from $\Psi_1$ we obtain the Klein-invariant combination
\be
\label{KleinInvariant1}
\hat{\Psi}_1 = \left(\begin{array}{c} \psi_1(x) \\ e^{i \alpha} \psi_1(x) \end{array}\right) e^{i k_4 x^4} e^{i k_5 x^5}
+ (-1)^{n_5} \left(\begin{array}{c} e^{i \alpha} i \bar{\gamma} \psi_1(x) \\ - i \bar{\gamma} \psi_1(x) \end{array}\right) e^{-i k_4 x^4} e^{i k_5 x^5}\,.
\ee
Starting from $\Psi_2$ we obtain what looks like a second Klein-invariant combination.
\be
\label{KleinInvariant2}
\hat{\Psi}_2 = \left(\begin{array}{c} \bar{\gamma} \psi_2(x) \\ - e^{i \alpha} \bar{\gamma} \psi_2(x) \end{array}\right) e^{i k_4 x^4} e^{i k_5 x^5}
+ (-1)^{n_5} \left(\begin{array}{c} -i e^{i \alpha} i \psi_2(x) \\ - i \psi_2(x) \end{array}\right) e^{-i k_4 x^4} e^{i k_5 x^5}\,.
\ee
However, this second set of solutions is redundant, since\footnote{To verify this, note that $\alpha$ changes sign under $n_4 \rightarrow - n_4$.}
\be
i e^{-i\alpha} (-1)^{n_5} \hat{\Psi}_2\big\vert_{n_4,n_5} = \hat{\Psi}_1\big\vert_{-n_4,n_5}\,.
\ee
That is, the two sets of modes are related by $n_4 \rightarrow - n_4$.  Overall, we end up with the complete set of linearly-independent modes given in (\ref{pin+modes}).

The breaking of $P$ and $C$ can be seen in the vev of the fermion bilinear (\ref{R4+bilinear}).  It can also be seen directly in the modes.  Acting on the modes (\ref{pin+modes}) with the parity transformation $P$ defined in (\ref{6Dparity}), it is
straightforward to see that
\bea
\nonumber
&& { P} \Psi(x,x^4,x^5) = \left\lbrace\begin{array}{l}
\left(\begin{array}{c} { p} \psi(x) \\ + i \bar{\gamma} { p} \psi(x) \end{array}\right) \qquad \hbox{\rm for $n_4 = n_5 = 0$} \\[20pt]
\left(\begin{array}{c} { p} \psi(x) \\ e^{i \alpha} { p} \psi(x) \end{array}\right) e^{i k_4 x^4} e^{i k_5 x^5}
- (-1)^{n_5} \left(\begin{array}{c} e^{i \alpha} i \bar{\gamma} { p} \psi(x) \\ - i \bar{\gamma} { p} \psi(x) \end{array}\right) e^{-i k_4 x^4} e^{i k_5 x^5}
\end{array}\right. \\
\label{Ppin+mode}
&& \hspace{7.1cm} \hbox{\rm for $(n_4,n_5) \not= (0,0)$.}
\eea
Here ${ p}$ is the 4D parity transformation defined in (\ref{4Dparity}).  Comparing (\ref{Ppin+mode}) to (\ref{pin+modes}), some crucial signs changed, so that
instead of satisfying the Klein bottle boundary conditions, the parity-transformed modes obey
${ P} \Psi(x) = - ({ P} \Psi)^K(x)$.  Note that, although the Klein bottle
breaks parity in 4D, we do not get chiral fermions in 4D as a result.  In this respect the Klein bottle cannot
provide an origin for the fermions of the standard model, even though it may be a source for $P$ violation.

For ${ R}_4^+$ boundary conditions, a similar result holds for $C$: acting on the modes (\ref{pin+modes}) with
charge conjugation ${ C}$ defined in (\ref{C}), it is straightforward to check that ${ C} \Psi(x) = - ({ C} \Psi)^K(x)$.
In this way parity as well as charge conjugation are broken by ${ R}_4^+$ boundary conditions, but the combination $CP$ is preserved.

\subsection{${ R}_4^-$ spectrum\label{sect:pin-KK}}
Finally, we consider the Kaluza-Klein spectrum with ${ R}_4^-$ boundary conditions.  There are a few minor but crucial changes compared to what we encountered
in section \ref{sect:pin+KK}.

We can still expand the field in momentum modes on the covering torus, with the momentum in $x^5$ shifted by half
a unit to make the field anti-periodic.
\bea
\nonumber
&& \Psi(x,x^4,x^5) = \left(\begin{array}{l}\chi_1(x) \\ \chi_2(x) \end{array}\right) e^{i k_4 x^4} e^{i k_5 x^5} \\[5pt]
&& k_4 = {n_4 \over r_4} \qquad k_5 = {n_5 \over 2 r_5} \qquad n_4 \in {\amsbb Z}, \, n_5 \in {\amsbb Z} + {1 \over 2}
\eea
The notation $(n_4,n_5)$ is consistent with the rest of the paper.  In this section it is convenient to set
$n_4 = \ell_4$, $n_5 = \ell_5 + {1 \over 2}$ so that modes are labeled by two integers $(\ell_4,\ell_5)$ with
\be
k_4 = {\ell_4 \over r_4} \qquad k_5 = {2 \ell_5 + 1 \over 4 r_5} \qquad \ell_4,\ell_5 \in {\amsbb Z}\,.
\ee
Note that there is no zero mode to worry about.  As before, we set $k_5 + i k_4 = m e^{-i \alpha}$.

The Dirac equation is unchanged, so just as in (\ref{CoveringModes1}), (\ref{CoveringModes2}) a complete set of modes on the covering torus is given by
\bea
\label{CoveringModes3}
&& \Psi_1(x,x^4,x^5) = \left(\begin{array}{c} \psi_1(x) \\ e^{i \alpha} \psi_1(x) \end{array}\right) e^{i k_4 x^4} e^{i k_5 x^5} \\
\label{CoveringModes4}
&& \Psi_2(x,x^4,x^5) = \left(\begin{array}{c} \bar{\gamma} \psi_2(x) \\ - e^{i \alpha} \bar{\gamma} \psi_2(x) \end{array}\right) e^{i k_4 x^4} e^{i k_5 x^5}\,.
\eea
Here $\psi_1$ and $\psi_2$ are massive Dirac fields in 4D, with
\be
i \gamma^\mu \partial_\mu \psi_1(x) = m \psi_1(x) \qquad i \gamma^\mu \partial_\mu \psi_2(x) = m \psi_2(x)\,.
\ee
Next we impose the Klein bottle boundary conditions, $\Psi(x) = \Psi^K(x)$ where
\be
\Psi^K(x,x^4,x^5) = i \Gamma^4 \bar{\Gamma} \Psi(x,-x^4,x^5 + 2 \pi r_5)\,.
\ee
In building the invariant combination $\Psi + \Psi^K$, note that $\Psi^K$ picks up an explicit factor of $i$ from the modified reflection operation $i \Gamma^4 \bar{\Gamma}$.
It picks up another factor of $i$ because shifting $x^5 \rightarrow x^5 + 2 \pi r_5$ on the covering torus produces a factor $i (-1)^{\ell_5}$ due to anti-periodicity of the modes
(\ref{CoveringModes3}), (\ref{CoveringModes4}).  Overall, compared to what we found in section \ref{sect:pin+KK}, there is an additional minus sign in the Klein-transformed field.
We can build an invariant combination starting from $\Psi_1$ or $\Psi_2$, but as in section \ref{sect:pin+KK} the second combination is redundant.\footnote{The relation is
$- i e^{-i\alpha} (-1)^{\ell_5} \big(\Psi_2 + \Psi_2^K\big)\big\vert_{\ell_4,\ell_5} = \big(\Psi_1 + \Psi_1^K\big)\big\vert_{-\ell_4,\ell_5}$.}

To summarize, with ${ R}_4^-$ boundary conditions the Kaluza-Klein tower is labeled by momenta
\be
k_4 = {\ell_4 \over r_4} \qquad k_5 = {2 \ell_5 + 1 \over 4 r_5} \qquad \ell_4,\ell_5 \in {\amsbb Z}\,.
\ee
For each Kaluza-Klein momentum there is a Dirac fermion $\psi(x)$ in 4D, with mass
\be
m = \sqrt{k_4^2 + k_5^2}\,.
\ee
The 4D field lifts to a solution to the 6D Dirac equation
\be
\label{pin-modes}
\Psi(x,x^4,x^5) = 
\left(\begin{array}{c} \psi(x) \\ e^{i \alpha} \psi(x) \end{array}\right) e^{i k_4 x^4} e^{i k_5 x^5}
- (-1)^{\ell_5} \left(\begin{array}{c} e^{i \alpha} i \bar{\gamma} \psi(x) \\ - i \bar{\gamma} \psi(x) \end{array}\right) e^{-i k_4 x^4} e^{i k_5 x^5}
\ee
where the phase $\alpha$ is defined by $k_5 + i k_4 = m e^{-i \alpha}$.

The ${ R}_4^-$ modes (\ref{pin-modes}) look almost identical to the ${ R}_4^+$ modes (\ref{pin+modes}).\footnote{We could make them look completely identical if we had
built the ${\rm pin}^-$ structure using $-i \Gamma^4 \bar{\Gamma}$.}  The difference is hidden in the shifted quantization condition for $k_5$.  What are the consequences?
For parity there is no consequence: $P$ is still broken, for reasons given at the end of section \ref{sect:pin+KK}.  What about charge conjugation?  This acts on the field by
\be
{ C} \, : \, \Psi(x) \rightarrow \Psi'(x) = \Gamma^2 \Gamma^4 \bar{\Gamma} \Psi^*(x) = \left(\begin{array}{cc} 0 & i \gamma^2 \bar{\gamma} \\ i \gamma^2 \bar{\gamma} & 0 \end{array}\right)
\Psi^*(x)\,.
\ee
Complex conjugation takes $k_5 \rightarrow - k_5$ or equivalently $n_5 \rightarrow - n_5$.  For ${ R}_4^+$, the modes have a factor $(-1)^{n_5}$ which is invariant under this transformation.
But for ${ R}_4^-$, sending $k_5 \rightarrow - k_5$ means $\ell_5 \rightarrow - \ell_5 - 1$.  The ${ R}_4^-$ modes have a factor $(-1)^{\ell_5}$ which changes sign under this transformation.
Thus the shifted quantization condition on $k_5$ produces an additional minus sign that leaves the set of modes (\ref{pin-modes}) invariant.  In more detail, under charge conjugation, the
${ R}_4^-$ modes transform amongst themselves as
\be
\Psi'\big\vert_{\ell_4,\ell_5,\psi(x)} = i (-1)^{\ell_5} \Psi\big\vert_{\ell_4,-\ell_5-1,-i\gamma^2\psi^*(x)}\,.
\ee
The notation indicates that $k_4$ stays the same, $k_5$ changes sign, and $\psi(x)$ is replaced with the 4D
charge-conjugate field $-i \gamma^2 \psi^*(x)$.
In this way $C$ is preserved by ${ R}_4^-$ boundary conditions but broken by ${ R}_4^+$.  We already knew this from the interplay of $C$ with the boundary conditions; this shows how it happens at the level of modes.

\section{Klein bottle correlators\label{appendix:correlator}}
Our goal is to evaluate the two-point correlator $\langle \Psi(y) \bar{\Psi}(x) \rangle$ for a
Dirac field on ${\amsbb R}^{3,1} \times K_2$.  We consider both ${ R}_4^+$ and ${ R}_4^-$
boundary conditions, and for completeness we include a mass term in 6D.
As a by-product, we obtain correlators for massive scalar
and pseudo-scalar fields.  These results are used to evaluate
fermion bilinears in sections \ref{sect:R4+bilinear} and \ref{sect:R4-bilinear}, and Casimir energies in section \ref{sect:Casimir}.
Related calculations on ${\amsbb R}^{1,1} \times K_2$ may be found in \cite{DeWitt:1979dd,DeWitt-Morette:1990nrn}.

\subsection{Momentum representation\label{sect:momentum}}
A convenient starting point is the two-point correlator for a Dirac field on the covering torus, in which we identify
\be
x^4 \approx x^4 + 2 \pi r_4, \qquad x^5 \approx x^5 + 4 \pi r_5\,.
\ee
We wish to consider both periodic and anti-periodic boundary conditions in the $x^5$ direction.  We denote these
correlators by ($+$ for periodic, $-$ for anti-periodic)
\be
S_{T^2}^\pm = \langle 0 \vert \Psi(y) \bar{\Psi}(x) \vert 0 \rangle\,.
\ee
We can construct two-point correlators on a Klein bottle with ${ R}_4^\pm$ boundary
conditions, denoted $S_{K^2}^\pm$, by introducing a single image charge on the covering torus.  This leads to
\bea
\label{SK2+}
&& S_{K_2}^+(y \vert x) = S_{T^2}^+(y \vert x) + \Gamma^4 \bar{\Gamma} S_{T^2}^+(\tilde{y} \vert x) \\[5pt]
\label{SK2-}
&& S_{K_2}^-(y \vert x) = S_{T^2}^-(y \vert x) + i \Gamma^4 \bar{\Gamma} S_{T^2}^-(\tilde{y} \vert x)\,.
\eea
Recall that the Klein image point is $\tilde{y} = (y^\mu,-y^4,y^5 + 2 \pi r_5)$.

The correlator with periodic boundary conditions in $x^5$,
\be
S_{T^2}^+ = \langle 0 \vert \Psi(y) \bar{\Psi}(x) \vert 0 \rangle_{\hbox{ \small periodic}}
\ee
can be obtained from the standard Dirac propagator in 6D
by quantizing the momenta so that $n_4,n_5 \in {\amsbb Z}$.
\bea
\nonumber
S_{T^2}^+(y \vert x) & = & i \int {d^4 p \over (2\pi)^4} \, {1 \over 2 \pi r_4} \sum_{n_4 \in {\amsbb Z}}
{1 \over 4 \pi r_5} \sum_{n_5 \in {\amsbb Z}} e^{-i p \cdot (y - x)} e^{i n_4 (y^4 - x^4) / r_4} e^{i n_5 (y^5 - x^5) / 2 r_5} \\[3pt]
\label{ST2+}
& & \hspace{1cm} {\Gamma^\mu p_\mu - \Gamma^4 {n_4 \over r_4} - \Gamma^5 {n_5 \over 2 r_5} + m \over p^2 - (n_4/r_4)^2 - (n_5 / 2 r_5)^2 - m^2 + i 0^+}
\eea
The poles are handled with a Feynman prescription, so (\ref{ST2+}) is a time-ordered correlator.  The correlator with anti-periodic boundary conditions,
\be
S_{T^2}^- = \langle 0 \vert \Psi(y) \bar{\Psi}(x) \vert 0 \rangle_{\hbox{ \small anti-periodic}}
\ee
is given by shifting the quantization condition so that $n_5 \in {\amsbb Z} + {1 \over 2}$.
\bea
\nonumber
S_{T^2}^-(y \vert x) & = & i \int {d^4 p \over (2\pi)^4} \, {1 \over 2 \pi r_4} \sum_{n_4 \in {\amsbb Z}}
{1 \over 4 \pi r_5} \sum_{n_5 \in {\amsbb Z} + {1 \over 2}} e^{-i p \cdot (y - x)} e^{i n_4 (y^4 - x^4) / r_4} e^{i n_5 (y^5 - x^5) / 2 r_5} \\
\label{ST2-}
& & \hspace{1cm} {\Gamma^\mu p_\mu - \Gamma^4 {n_4 \over r_4} - \Gamma^5 {n_5 \over 2 r_5} + m \over p^2 - (n_4/r_4)^2 - (n_5 / 2 r_5)^2 - m^2 + i 0^+}
\eea

In principle we are done, but for practical purposes it is convenient to put the torus correlators (\ref{ST2+}), (\ref{ST2-}) in a more useful form.
For simplicity we work at equal times, $y^0 = x^0$, and later restore 4D Lorentz invariance.  We Wick rotate the momentum
\be
p_E^\mu = (-i p^0,{\bf p}) \qquad\quad \int d^4p = i \int d^4p_E
\ee
and introduce a Schwinger parameter $s$ to represent
\be
{1 \over p_E^2 + m^2} = \int_0^\infty ds \, e^{-s(p_E^2 + m^2)}\,.
\ee
Performing the Gaussian integral over 4D momentum, restoring 4D Lorentz invariance, and introducing a proper-time cutoff $\epsilon \rightarrow 0$
as a UV regulator, the momentum sums can be expressed in terms of Jacobi theta functions.
\bea
\nonumber
S_{T^2}^\pm(y \vert x) & = & \int_{\epsilon^2}^\infty {ds \over 16 \pi^2 s^2} \, e^{-sm^2} \left(i \Gamma^M {\partial \over \partial y^M} + m\right)
e^{\big((y - x)^2 - i 0^+\big)/4s} \\
\label{ST2pm}
& & {1 \over 2 \pi r_4} \theta_3\Big({y^4 - x^4 \over 2 r_4},e^{-s/r_4^2}\Big)
{1 \over 4 \pi r_5} \theta_{\scriptstyle 3 \atop \scriptstyle 2}\Big({y^5 - x^5 \over 4 r_5},e^{-s/(2 r_5)^2}\Big)
\eea
The correlator with periodic boundary conditions $S_{T^2}^+$ has a Jacobi $\theta_3$ function in the last factor, while the correlator with
anti-periodic boundary conditions
$S_{T^2}^-$ has $\theta_2$.  The Feynman prescription can be thought of as giving $y^0$ a small negative imaginary part if $y^0 > x^0$,
and a small positive imaginary part if $y^0 < x^0$.\footnote{This yields a time-ordered correlator.  See for example \cite{Birrell:1982ix}, pages 23 and 76.}
To obtain the correlator on a Klein bottle, simply substitute (\ref{ST2pm}) into (\ref{SK2+}), (\ref{SK2-}).

\subsection{Theta function identities\label{sect:theta}}
We pause to record a few useful properties of the Jacobi theta functions \cite{NIST:DLMF}.  For the $x^4$ direction we have
\bea
\label{theta3x4n}
{1 \over 2 \pi r_4} \theta_3\Big({y^4 - x^4 \over 2 r_4},e^{-s/r_4^2}\Big) & = &
{1 \over 2 \pi r_4} \sum_{n_4 \in {\amsbb Z}} e^{-s n_4^2/r_4^2} e^{i n_4 (y^4 - x^4)/r_4} \\
\label{theta3x4w}
& = & {1 \over \sqrt{4 \pi s}} \sum_{w_4 \in {\amsbb Z}} e^{-(y^4 - x^4 + 2 \pi r_4 w_4)^2/4s} \,.
\eea
The first line expresses $\theta_3$ as a momentum sum.  The second line, obtained by a modular transformation (or equivalently
by Poisson resummation), expresses $\theta_3$ as a winding sum or equivalently as a sum over image charges.

With periodic boundary conditions in $x^5$, the relevant theta function can be obtained from the above by replacing $r_4 \rightarrow 2 r_5$.
\bea
\label{theta3x5n}
{1 \over 4 \pi r_5} \theta_3\Big({y^5 - x^5 \over 4 r_5},e^{-s/(2r_5)^2}\Big) & = &
{1 \over 4 \pi r_5} \sum_{n_5 \in {\amsbb Z}} e^{-s n_5^2/(2r_5)^2} e^{i n_5 (y^5 - x^5)/2r_5} \\
\label{theta3x5w}
& = & {1 \over \sqrt{4 \pi s}} \sum_{w_5 \in {\amsbb Z}} e^{-(y^5 - x^5 + 4 \pi r_5 w_5)^2/4s} 
\eea
For anti-periodic boundary conditions in $x^5$ the relevant theta function is
\bea
{1 \over 4 \pi r_5} \theta_2\Big({y^5 - x^5 \over 4 r_5},e^{-s/(2r_5)^2}\Big) & = &
{1 \over 4 \pi r_5} \sum_{n_5 \in {\amsbb Z} + {1 \over 2}} e^{-s n_5^2/(2r_5)^2} e^{i n_5 (y^5 - x^5)/2 r_5} \\
\label{theta2x5w}
& = & {1 \over \sqrt{4 \pi s}} \sum_{w_5 \in {\amsbb Z}} (-1)^{w_5} e^{-(y^5 - x^5 + 4 \pi r_5 w_5)^2/4s} \,.
\eea

\subsection{Image charge representation\label{sect:image}}
The theta function identities of section \ref{sect:theta} enable us to re-write the torus correlator (\ref{ST2pm}) as a sum over
image charges.
\bea
\nonumber
S_{T^2}^\pm(y \vert x) & = &  \left(i \Gamma^M {\partial \over \partial y^M} + m\right) \int_{\epsilon^2}^\infty {ds \over 64 \pi^3 s^3} \, e^{-sm^2}
\sum_{w_4,\,w_5 \in {\amsbb Z}} \left\lbrace {1 \atop (-1)^{w_5}} \right\rbrace \\
& & \quad e^{\big((y - x)^2 - (y^4 - x^4 + 2 \pi r_4 w_4)^2 - (y^5 - x^5 + 4 \pi r_5 w_5)^2 - i 0^+\big)/4s}
\eea
The factor in curly braces stands for $1$ with ${ R}_4^+$ boundary conditions, and for $(-1)^{w_5}$ with ${ R}_4^-$.  In the massless limit
the proper-time integral becomes elementary and, dropping the UV regulator, the result for $S_{T^2}^\pm$ simplifies to
\be
i\Gamma^M {\partial \over \partial y^M} \sum_{w_4,\,w_5 \in {\amsbb Z}} \left\lbrace {1 \atop (-1)^{w_5}} \right\rbrace
{1 \over 4 \pi^3 \left[(y - x)^2 - (y^4 - x^4 + 2 \pi r_4 w_4)^2 - (y^5 - x^5 + 4 \pi r_5 w_5)^2 - i 0^+\right]^2}\,.
\ee
Incidentally, erasing $i \Gamma^M {\partial \over \partial y^M}$ produces an image-charge representation of the massless
scalar propagator that we will encounter in the next section.

\subsection{Scalar correlator\label{sect:scalar}}
Having obtained the correlator for a Dirac field, it is straightforward to extract the correlator for a scalar field $\phi$.
We consider both scalar and pseudoscalar fields.  These are characterized by periodicity in $x^4$,
\be
\phi(x, x^4, x^5) = \phi(x,x^4 + 2 \pi r_4, x^5)\,,
\ee
and periodicity twisted by parity in $x^5$,
\be
\label{scalarbc}
\begin{array}{ll}
\hbox{\rm scalar:} & \phi(x,x^4, x^5) = \phi(x,-x^4, x^5 + 2 \pi r_5) \\[5pt]
\hbox{\rm pseudoscalar:} & \phi(x,x^4, x^5) = - \phi(x,-x^4, x^5 + 2 \pi r_5)\,.\end{array}
\ee
Note that in both cases, the scalar field is periodic on the covering torus.

The scalar correlator on the covering torus with periodic boundary conditions, denoted $D_{T^2}(y \vert x)$, can be obtained from the Dirac propagator (\ref{ST2+})
by erasing a factor of $\Gamma^M p_M + m$.  Following this change through, we find
\bea
\nonumber
D_{T^2}(y \vert x) & = & \int_{\epsilon^2}^\infty {ds \over 16 \pi^2 s^2} \, e^{-sm^2} e^{\big((y - x)^2 - i 0^+\big)/4s} \\
\label{DT2} & &
{1 \over 2 \pi r_4} \theta_3\Big({y^4 - x^4 \over 2 r_4},e^{-s/r_4^2}\Big)
{1 \over 4 \pi r_5} \theta_3\Big({y^5 - x^5 \over 4 r_5},e^{-s/(2 r_5)^2}\Big)\,.
\eea
The scalar and pseudoscalar correlators on a Klein bottle, denoted $D_{K_2}^\pm$, are given by introducing an image charge to enforce the boundary conditions (\ref{scalarbc}).
\be
\label{DK2}
\begin{array}{ll}
\hbox{\rm scalar:} & D_{K^2}^+(y \vert x) = D_{T^2}(y \vert x) + D_{T^2}(\tilde{y} \vert x) \\[5pt]
\hbox{\rm pseudoscalar:} & D_{K^2}^-(y \vert x) = D_{T^2}(y \vert x) - D_{T^2}(\tilde{y} \vert x)
\end{array}
\ee

\section{${ R}_4^\theta$ boundary conditions and symmetries\label{appendix:pinC}}
Here we consider ${ R}_4^\theta$ boundary conditions (\ref{R4theta}), in which the Klein bottle is constructed using a reflection ${ R}_4$ with a phase $e^{i \theta/2}$.
\bea
\nonumber
&& \Psi(x^\mu,x^4,x^5) = \Psi(x^\mu,x^4 + 2 \pi r_4,x^5) \\[2pt]
\label{R4theta2}
&& \Psi(x^\mu,x^4,x^5) = e^{i \theta / 2} { R}_4 \Psi(\tilde{x}) = e^{i \theta / 2} \Gamma^4 \bar{\Gamma} \Psi(x^\mu,-x^4,x^5 + 2 \pi r_5)
\eea
This defines a ${\rm pin}_{\amsbb C}$ structure on the Klein bottle.  Going twice around the Klein bottle, we have
\be
\label{R4thetaphase}
\Psi(x^\mu,x^4,x^5) = e^{i \theta / 2} { R}_4 \Psi(\tilde{x}) = e^{i \theta} \Psi(x^\mu, x^4, x^5 + 4 \pi r_5)\,.
\ee
So the boundary conditions on the covering torus are twisted by an arbitrary phase.  This interpolates between ${ R}_4^+$ at $\theta = 0$ and ${ R}_4^-$ at $\theta = \pi$.

It is straightforward to determine the fate of the discrete symmetries ${ P}$, ${ R}_4$, ${ R}_5$, ${ C}$ given these boundary conditions.  For example,
consider the internal reflection on the Klein bottle ${ R}_5$, defined by
\be
{ R}_5: \!\quad \Psi'(x) = \Gamma^5 \bar{\Gamma} \Psi(x^\mu,x^4,-x^5)\,.
\ee
Suppose $\Psi$ satisfies the ${ R}_4^\theta$ boundary conditions.  Does $\Psi'$ also satisfy the boundary conditions?  We have
\bea
\nonumber
\Psi'(x) & = & \Gamma^5 \bar{\Gamma} \Psi(x^\mu,x^4,-x^5) \\
\nonumber
& = & \Gamma^5 \bar{\Gamma} e^{i \theta / 2} \Gamma^4 \bar{\Gamma} \Psi(x^\mu,-x^4,-x^5 + 2 \pi r_5) \\
& = & \Gamma^5 \bar{\Gamma} e^{i \theta / 2} \Gamma^4 \bar{\Gamma} e^{-i \theta} \Psi(x^\mu,-x^4,-x^5 - 2 \pi r_5) \\
\nonumber
& = & \Gamma^5 \bar{\Gamma} e^{i \theta / 2} \Gamma^4 \bar{\Gamma} e^{-i \theta} \Gamma^5 \bar{\Gamma} \Psi'(x^\mu,-x^4,x^5 + 2 \pi r_5)\,.
\eea
In the first line we used the definition of $\Psi'$, in the second we used the ${ R}_4^\theta$ boundary conditions, and in the third we used the boundary
conditions (\ref{R4thetaphase}) on the covering torus.  In the last line we used the definition of $\Psi'$ in reverse.  After a bit of Dirac algebra, this becomes
\be
\Psi'(x) = - e^{-i \theta} e^{i \theta / 2} \Gamma^4 \bar{\Gamma} \Psi'(\tilde{x})\,.
\ee
Compared to (\ref{R4theta2}), there is an extra phase $- e^{-i \theta}$ on the right hand side.  We have recorded this under ${ R}_5$ in Table \ref{table:symmetries}.
The fates of the other symmetries can be worked out along similar lines, and are listed in the table.

As promised, the ${ R}_4^\theta$ boundary conditions interpolate from ${ R}_4^+$ at $\theta = 0$ to ${ R}_4^-$ at $\theta = \pi$.  For generic values of
$\theta$, the only unbroken symmetry is the internal reflection ${ R}_4$.  At $\theta = 0$ and $\theta = \pi$ there are additional unbroken symmetries, listed in
sections \ref{sect:R4+symmetry} and \ref{sect:R4-symmetry}.  There are two other special points.  At $\theta = \pi/2$ and $\theta = 3\pi/2$, the combination
${ C R}_5$ flips the sign of the boundary conditions.  At these values of $\theta$ there is an enhanced symmetry, with ${ CPR}_5$ and ${ CPR}_4{ R}_5$ unbroken.

\section{${ CR}_4^+$ boundary conditions and symmetries\label{appendix:CR}}
Here we consider the ${ CR}_4^+$ boundary condition (\ref{CR4condition}), in which the Klein bottle is constructed using a twist by parity combined
with charge conjugation.
\bea
\nonumber
&& \Psi(x^\mu,x^4,x^5) = \Psi(x^\mu,x^4 + 2 \pi r_4,x^5) \\[2pt]
&& \Psi(x^\mu,x^4,x^5) = -i \Gamma^2 \Psi^*(x^\mu,-x^4,x^5+2\pi r_5)
\eea
We are using the matrix $-i { CR}_4 = - i \Gamma^2$ to produce the twist.  The factor of $-i$ in the transformation is a
convenient but somewhat arbitrary choice of phase, which makes $-i \Gamma^2$ real.  Since $\left(-i\Gamma^2\right)^2 = \identity$,
this defines a ${\rm pin}^+$ structure on the Klein bottle.\footnote{We have not found boundary conditions that involve charge
conjugation and define a ${\rm pin}^-$ structure.}  Going twice around the Klein bottle, we have
\be
\Psi(x^\mu,x^4,x^5) = -i \Gamma^2 \Psi^*(x^\mu,-x^4,x^5+2\pi r_5) = \Psi(x^\mu,x^4,x^5 + 4 \pi r_5)\,.
\ee
So the field is periodic on the covering torus.

Just as in section \ref{sect:R4+}, the ${ CR}_4^+$ boundary conditions break translation invariance.  More accurately translations
in the $x^4$ direction
\be
\Psi'(x) = \Psi(x^\mu,x^4 + a, x^5)
\ee
are broken to the ${\amsbb Z}_2$ subgroup $a \in \lbrace 0, \pi r_4 \rbrace$.  However, unlike the boundary conditions studied
previously, ${ CR}_4^+$ preserves chiral symmetry.  That is, if $\Psi(x)$ satisfies the ${ CR}_4^+$ boundary conditions, then
so does
\be
\Psi'(x) = e^{i \alpha \bar{\Gamma}} \Psi(x) \qquad \hbox{\rm for any $e^{i \alpha} \in U(1)$}\,.
\ee

The fate of the discrete symmetries $P$, ${ R}_4$, ${ R}_5$, $C$ can be studied just as in sections \ref{sect:R4+}
and \ref{sect:R4-}.  Recall that we define these transformations as follows.
\bea
\label{CR-P}
&& { P} : \,\quad \Psi'(x) =  \left(\begin{array}{cc} \gamma^0 & 0 \\ 0 & \gamma^0 \end{array}\right) \Psi(t,-{\bf x},x^4,x^5) \\[3pt]
\label{CR-R4}
&& { R}_4 : \!\quad \Psi'(x) = \Gamma^4 \bar{\Gamma} \Psi(x^\mu,-x^4,x^5) \\[3pt]
\label{CR-R5}
&& { R}_5: \!\quad \Psi'(x) = \Gamma^5 \bar{\Gamma} \Psi(x^\mu,x^4,-x^5) \\[3pt]
\label{CR-C}
&& { C} : \,\quad \Psi'(x) = \Gamma^2 \Gamma^4 \bar{\Gamma} \Psi^*(x)
\eea
It is straightforward to check that $P$ and ${ R}_4$ are violated by the boundary conditions, while ${ R}_5$ and $C$
are preserved.  For example, if $\Psi(x)$ satisfies the ${ CR}_4^+$ boundary conditions, one can show that the parity-transformed
field $\Psi'(x)$ defined in (\ref{CR-P}) obeys
\be
\label{CR4+Psi'}
\Psi'(x) = + i \Gamma^2 \big(\Psi')^*(\tilde{x})
\ee
with a $+$ sign rather than a $-$ sign on the right hand side.  So the parity-transformed field violates the boundary conditions.

The transformations (\ref{CR-P}) -- (\ref{CR-C}) can be generalized by introducing a phase $e^{i \phi / 2}$ on the right-hand side.  It is straightforward to carry this phase through
and show that it appears in the boundary conditions as $e^{i \phi}$.  For example, (\ref{CR4+Psi'}) gains a factor $e^{i \phi}$ on the right-hand side.
We have recorded these results in Table \ref{table:symmetries}.  Absent further restrictions,
we are free to choose the phase for $C$ independently of the phase used to define $P$.  It is then possible to choose phases
so that both $P$ and $C$ are preserved by ${ CR}_4^+$ boundary conditions.

With ${ CR}_4^+$ boundary conditions, it does not seem possible to move away from a ${\rm pin}^+$ structure, although as we have just seen it is meaningful to introduce a
phase in the would-be symmetry transformations.  With ${ R}_4^+$ the situation is reversed.  The boundary conditions can be generalized, to ${ R}_4^\theta$, but introducing a phase
in the would-be symmetry transformations has no effect on whether or not they are broken.


\begin{thebibliography}{10}

\bibitem{Dabholkar:1996pc}
A.~Dabholkar and J.~Park, ``{Strings on orientifolds},''
  \href{http://dx.doi.org/10.1016/0550-3213(96)00395-1}{{\em Nucl. Phys. B}
  {\bfseries 477} (1996) 701--714},
  \href{http://arxiv.org/abs/hep-th/9604178}{{\ttfamily arXiv:hep-th/9604178}}.

\bibitem{Freed:2019sco}
D.~S. Freed and M.~J. Hopkins, ``{Consistency of M-theory on non-orientable
  manifolds},'' \href{http://dx.doi.org/10.1093/qmath/haab007}{{\em The
  Quarterly Journal of Mathematics} {\bfseries 72} no.~1-2, (2021) 603--671},
  \href{http://arxiv.org/abs/1908.09916}{{\ttfamily arXiv:1908.09916
  [hep-th]}}.

\bibitem{Cheng:2023owv}
P.~Cheng, I.~V. Melnikov, and R.~Minasian, ``{Flat F-theory and friends},''
  \href{http://dx.doi.org/10.1007/JHEP01(2024)027}{{\em JHEP} {\bfseries 01}
  (2024) 027}, \href{http://arxiv.org/abs/2306.00865}{{\ttfamily
  arXiv:2306.00865 [hep-th]}}.

\bibitem{Chakrabhavi:2025bfi}
V.~Chakrabhavi, A.~Debray, M.~Dierigl, and J.~J. Heckman, ``{Exploring
  pintopia: Reflection branes, bordisms, and U-dualities},''
  \href{http://arxiv.org/abs/2509.03573}{{\ttfamily arXiv:2509.03573
  [hep-th]}}.

\bibitem{Appelquist:1987nr}
T.~Appelquist, A.~Chodos, and P.~G.~O. Freund, eds., {\em {Modern Kaluza-Klein
  theories}}.
\newblock Addison-Wesley, Reading, USA, 1987.

\bibitem{Witten:2015aba}
E.~Witten, ``{Fermion path integrals and topological phases},''
  \href{http://dx.doi.org/10.1103/RevModPhys.88.035001}{{\em Rev. Mod. Phys.}
  {\bfseries 88} no.~3, (2016) 035001},
  \href{http://arxiv.org/abs/1508.04715}{{\ttfamily arXiv:1508.04715
  [cond-mat.mes-hall]}}.

\bibitem{BERG_2001}
M.~Berg, C.~DeWitt-Morette, S.~Gwo, and E.~Kramer, ``The pin groups in physics:
  C, P, and T,'' \href{http://dx.doi.org/10.1142/s0129055x01000922}{{\em
  Reviews in Mathematical Physics} {\bfseries 13} no.~08, (Aug., 2001) 953},
  \href{http://arxiv.org/abs/0012006}{{\ttfamily arXiv:0012006 [math-ph]}}.

\bibitem{Freed:2016rqq}
D.~S. Freed and M.~J. Hopkins, ``{Reflection positivity and invertible
  topological phases},'' \href{http://dx.doi.org/10.2140/gt.2021.25.1165}{{\em
  Geom. Topol.} {\bfseries 25} (2021) 1165--1330},
  \href{http://arxiv.org/abs/1604.06527}{{\ttfamily arXiv:1604.06527
  [hep-th]}}.

\bibitem{DeWitt:1979dd}
B.~S. DeWitt, C.~F. Hart, and C.~J. Isham, ``{Topology and quantum field
  theory},'' \href{http://dx.doi.org/10.1016/0378-4371(79)90207-3}{{\em Physica
  A} {\bfseries 96} no.~1-2, (1979) 197--211}.

\bibitem{DeWitt-Morette:1990nrn}
C.~DeWitt-Morette and B.~S. DeWitt, ``{Pin groups in physics},''
  \href{http://dx.doi.org/10.1103/PhysRevD.41.1901}{{\em Phys. Rev. D}
  {\bfseries 41} (1990) 1901--1907}.

\bibitem{Peskin:1995ev}
M.~E. Peskin and D.~V. Schroeder,
  \href{http://dx.doi.org/10.1201/9780429503559}{{\em {An Introduction to
  quantum field theory}}}.
\newblock Addison-Wesley, Reading, USA, 1995.

\bibitem{Polchinski:1998rq}
J.~Polchinski, \href{http://dx.doi.org/10.1017/CBO9780511816079}{{\em {String
  theory. Vol. 1: An introduction to the bosonic string}}}.
\newblock Cambridge Monographs on Mathematical Physics. Cambridge University
  Press, 12, 2007.

\bibitem{Polchinski:1998rr}
J.~Polchinski, \href{http://dx.doi.org/10.1017/CBO9780511618123}{{\em {String
  theory. Vol. 2: Superstring theory and beyond}}}.
\newblock Cambridge Monographs on Mathematical Physics. Cambridge University
  Press, 12, 2007.

\bibitem{Hosotani:1988bm}
Y.~Hosotani, ``{Dynamics of nonintegrable phases and gauge symmetry
  breaking},'' \href{http://dx.doi.org/10.1016/0003-4916(89)90015-8}{{\em
  Annals Phys.} {\bfseries 190} (1989) 233}.

\bibitem{Scherk:1978ta}
J.~Scherk and J.~H. Schwarz, ``{Spontaneous breaking of supersymmetry through
  dimensional reduction},''
  \href{http://dx.doi.org/10.1016/0370-2693(79)90425-8}{{\em Phys. Lett. B}
  {\bfseries 82} (1979) 60--64}.

\bibitem{Birrell:1982ix}
N.~D. Birrell and P.~C.~W. Davies,
  \href{http://dx.doi.org/10.1017/CBO9780511622632}{{\em {Quantum Fields in
  Curved Space}}}.
\newblock Cambridge Monographs on Mathematical Physics. Cambridge Univ. Press,
  Cambridge, UK, 2, 1984.

\bibitem{NIST:DLMF}
``{\it NIST Digital Library of Mathematical Functions}.''
\newblock \url{https://dlmf.nist.gov/}. F.~W.~J. Olver, A.~B. {Olde Daalhuis},
  D.~W. Lozier, B.~I. Schneider, R.~F. Boisvert, C.~W. Clark, B.~R. Miller,
  B.~V. Saunders, H.~S. Cohl, and M.~A. McClain, eds. See sections
  \href{https://dlmf.nist.gov/20.2}{20.2} and
  \href{https://dlmf.nist.gov/20.7}{20.7}.

\end{thebibliography}
\providecommand{\href}[2]{#2}\begingroup\raggedright\endgroup

\end{document}